\documentclass[a4paper,fleqn,usenatbib]{mnras}

\usepackage[T1]{fontenc}
\usepackage{ae,aecompl}

\usepackage{graphicx}	
\usepackage{amsmath}	
\usepackage{amssymb}	
\usepackage{upgreek}
\usepackage{braket}
\usepackage{siunitx}
\usepackage{color}
\usepackage[dvipsnames]{xcolor}
\usepackage{footmisc}

\title[GRMHD thin discs]{Jets, disc-winds and oscillations in general relativistic, magnetically driven flows around black hole}

\author[Dihingia et al.]{
Indu K. Dihingia ,$^{1}$\thanks{E-mail: idihingia@iiti.ac.in}
Bhargav Vaidya ,$^{1}$
and Christian Fendt$^{2}$
\\
$^{1}$Department of Astronomy, Astrophysics and Space Engineering, 
                 Indian Institute of Technology Indore, Khandwa Road, Simrol, 453552, India\\
$^{2}$Max Planck Institute for Astronomy, K\"onigstuhl 17, DE-69117 Heidelberg, Germany\\
}

\date{Accepted XXX. Received YYY; in original form ZZZ}

\pubyear{2016}

\begin{document}
\label{firstpage}
\pagerange{\pageref{firstpage}--\pageref{lastpage}}
\maketitle

\begin{abstract}
Relativistic jets and disc-winds are typically observed in BH-XRBs and AGNs. 
However, many physical details of jet launching and the driving of disc winds from the underlying accretion disc are still not fully understood. 
In this study, we further investigate the role of the magnetic field strength and structure in launching jets and disc winds. 
In particular, we explore the connection between jet, wind, and the accretion disc around the central black hole. 
We perform axisymmetric GRMHD simulations of the accretion-ejection system using adaptive mesh refinement. 
Essentially, our simulations are initiated with a thin accretion disc in equilibrium. 
An extensive parametric study by choosing different combinations of magnetic field strength and initial magnetic field inclination is also performed. 
Our study finds relativistic jets driven by the Blandford \& Znajek (BZ) mechanism and the disc-wind driven by the Blandford \& Payne (BP) mechanism. 
We also find that plasmoids are formed due to the reconnection events, and these plasmoids advect with disc-winds. 
As a result, the tension force due to the poloidal magnetic field is enhanced in the inner part of the accretion disc, resulting in disc truncation and oscillation. 
These oscillations result in flaring activities in the jet mass flow rates. 
We find simulation runs with a lower value of the plasma-$\beta$, and lower inclination angle parameters are more prone to the formation of plasmoids and subsequent inner disc oscillations. Our models provide a possible template to understand spectral state transition phenomena in BH-XRBs.
\end{abstract}

\begin{keywords}
accretion, accretion discs - black hole physics - GRMHD - relativistic jets - disc wind - plasmoids
\end{keywords}

\section{Introduction}
Magnetic fields are ubiquitous in astrophysical environments and play a vital role in driving accretion flow around the black hole. 
With the increase of the sophistication of astrophysical observations, developing an unified numerical model for accretion flow is essential. General relativistic magneto-hydrodynamical (GRMHD) simulations have been instrumental for modeling accretion flow around the black hole \citep{Balbus-Hawley1998, Han2017}.
The existing numerical simulation codes are becoming mature day by day \citep{Porth-etal2019}; there are still huge discrepancies between theoretical predictions and observational findings for Active galactic nuclei (AGNs) as well as black hole X-ray binaries (BH-XRBs) \citep{Davis-Tchekhovskoy2020}. 

Astrophysical jets can be seen in young stellar objects, BH-XRBs, gamma-ray bursts (GRBs), and AGNs (see \cite{Fernandes-etal2012, Fender-Gallo2014, Davis-Tchekhovskoy2020} etc.). 
There are several studies related to the mechanism for energy extraction and jet launching from black holes. The Penrose process around the rotating black hole allows energy extraction from the infalling matter (see \citet{Misner-etal1973}). The generalized extension of the Penrose process, including the dynamo effects of a uniform magnetic field is known as the Blandford \& Znajeck (BZ) mechanism \citep{Blandford-Znajek1977, Komissarov2004}. 
In the past several decades, robust GRMHD codes have been developed to understand a number of aspects of the black hole and accretion process.  \cite{Koide-etal1999} discussed the possibility of the magnetically driven relativistic jet from the Schwarzschild space-time. \cite{DeVilliers-etal2003} studied the accretion processes in Kerr space-time and investigated the dependencies of accretion over the spin parameter. 
\cite{McKinney-Gammie2004} estimated outward energy flux from the Kerr black hole horizon to demonstrate the BZ mechanism. 
Further, \cite{Tchekhovskoy-etal2010, Tchekhovskoy-etal2011} measured the energy extracted by the BZ mechanism from the extremely rotating black holes using 2D as well as 3D GRMHD simulations. Additionally, these GRMHD codes have recently been augmented with radiative transfer processes to study the emission spectrum from the accretion disc \citep{Noble-etal2011, McKinney-etal2014,Dexter2016, Younsi-etal2016,Bronzwaer-etal2018, Bronzwaer-etal2020,Moscibrodzka2020, Chatterjee-etal2020}. 
Recent GRMHD simulations also show spontaneous jet formation from rapidly rotating black holes (e.g., \cite{Qian-etal2018, Vourellis-etal2019, Nathanail-etal2020}), which suggests that most of the energy extracted from the jet comes from the black hole rather than the accreting matter. Observations suggest that the radio luminosity of jet and black hole spin are correlated, and the jet power roughly increases as a square of the spin parameter \citep{Tchekhovskoy-etal2010, Tchekhovskoy-etal2011, Narayan-McClintock2012}. 
This property of the jet power is handy to determine the spin of the black hole using the continuum fitting method (see \citet{Zhang-etal1997, McClintock-etal2006, You-etal2016}).

\citet{Blandford-Payne1982} suggests that the matter can leave the surface of the accretion disc due to magneto-centrifugal acceleration, commonly known as Blandford \& Payne (BP) mechanism. Based on the current understanding, both BZ and BP mechanisms act simultaneously in launching the astrophysical jet (e.g., \cite{Hardee-etal2007, Xie-etal2012}. Due to very little observational evidences, it is impossible to pinpoint the dominating mechanism for jet launching. The VLBI observations suggest that the radius of the launching site of M87 is around $5.5 R_s$ (Schwarzschild radii) \citep{Doeleman-etal2012}, which essentially indicate that the BZ mechanism may be responsible for jet launching in M87. Similar observations for Cygnus A jet suggest the launching radius of the jet to be about $227 R_s$ \citep{Boccardi-etal2016}, hinting the presence of jet/wind driven through the BP mechanism.

Most of the astrophysical phenomena related to black holes are induced by the underlying accretion process. 
There are some extensive studies to understand the role of the spin parameters on the BZ process or the BP process (e.g., \cite{McKinney-etal2012, Vourellis-etal2019}). 
But, limited efforts are put into understanding the role of the disc flow parameters on energy extraction processes and subsequent launching of jets and winds. 
Few studies have shown that the initial magnetic field structure inside the torus plays a crucial role in the properties of the launched jets from the accretion disc \citep{Beckwith-etal2008,  Beckwith-etal2009, Nathanail-etal2020}. 
Motivated by these studies, we plan to study the role of the inclination angle of the initial magnetic fields on the jet and wind.  Subsequently, many steady analytic studies suggest that the strength of the magnetic field also plays a crucial role in deciding the radiative and dynamic properties of the accretion disc around black holes \citep[references therein]{Oda-etal2007, Oda-etal2012, Dihingia-etal2018, Dihingia-etal2020}. Here, we will also explore the role of the strength of the magnetic field in the dynamic properties of the jet and wind.

In this study, we simulate a thin accretion disc assuming axisymmetry and investigate its evolution features depending on the plasma-$\beta$ parameter and the magnetic field structure. 
The standard disc model is one of the most widely used thin accretion disc models in astrophysics, proposed by \citet{Shakura-Sunyaev1973}. In the same year, \citet[hereafter NT]{Novikov-Thorne1973} generalized the standard disc model to its general relativistic version. NT model is derived based on several assumptions. 
The most questionable assumption is that the $r-\phi$ component of the stress vanishes at the innermost stable circular orbit (ISCO) \citep{Thorne1974}. 
Several tests have been performed using magnetohydrodynamics (MHD) and general relativistic magnetohydrodynamics (GRMHD) simulations to understand these assumptions. 
MHD simulations using pseudo-Newtonian potential show variations of $\sim 10\%$ from the NT model \citep{Hawley-etal2002, Noble-etal2009}, but the GRMHD simulation shows the order of $\sim 3\%$ variation from the NT model \citep{Penna-etal2010}. We consider the flow profile along the equatorial plane of the thin accretion disc following the NT model in our initial setup.  The density profile at the off-equatorial plane is considered to be decaying exponentially with respect to the equatorial plane density \citep{Matsumoto-etal1984, Oda-etal2010}. We obtain the decaying length scale or disc height following \citet{Riffert-Herold1995, Peitz-Appl1997}, where the thin disc is in vertical equilibrium in general relativistic hydrodynamics framework.

Our paper is structured as follows. In section 2, we describe our model setup, including the basic equations for ideal GRMHD, essential assumptions we make, the initial conditions, and the parametric models. Subsequently, in sections 3, 4, 5, 6, and 7, we discuss our results from the different simulation models. In particular, we illustrate the launching of jets by the Blandford \& Znajek mechanism, the launching of disk winds. We further discuss their connection with the accretion disc and the corresponding oscillations of the inner part of the disc. In section 8, we encompass the possible implementation of our models in the astrophysical context. Finally, section 9 summarizes our work along with future outlooks. 

\section{Model setup}
In this section, we first discuss the governing equations of GRMHD (general relativistic magnetohydrodynamic) along  with our numerical setup and different simulation models adopted in this work.

\subsection{GRMHD equations}
The \texttt{code BHAC} \citep{Porth-etal2017, Olivares-etal2019} adopted for the present work solves the GRMHD 
equations, which constitute by the mass conservation, energy-momentum conservation, and Faraday's law. 
These equations are expressed in covariant notation as follows,
\begin{align}
\begin{aligned}
&\nabla_\mu\left(\rho u^\mu\right)=0,\\
&\nabla_\mu T^{\mu\nu}=0,\\
&\nabla_\mu{}^*F^{\mu\nu}=0,\\
\end{aligned}
\label{eq-01}
\end{align}
where $\rho$, $u^\mu$, $T^{\mu\nu}$, and ${}^*F^{\mu\nu}$ are the rest-mass density, the fluid four velocities, the energy-momentum tensor, and the dual of the Faraday tensor, respectively. 
The code is equipped with a fully adaptive mesh-refinement (AMR) framework. 
It utilizes the constrained-transport method \citep{DelZanna-etal2007} to maintain a divergence-free magnetic field throughout the simulation domain \citep{Olivares-etal2019}. The GRMHD equations are solved in a spherically symmetric, Modified Kerr-Schild (MKS) geometry. By adjusting the MKS stretching parameter, we concentrate maximum resolution around the equatorial plane in the domain of the simulation \citep{McKinney-Gammie2004}. 

In addition, we consider flow to be axi-symmetric, and therefore we only solve GRMHD equations in 2 spatial directions $(r, \theta)$. To do that, we use a logarithmic grid in the radial direction and implement the AMR grid to resolve the simulation domain of our interest efficiently. Further, we use a unit system with $G=M=c=1$, where $M$, $G$, and $c$ are the mass of the central black hole, gravitational constant, and speed of light. Subsequently, length, angular momentum, and time are expressed in terms of $GM/c^2$, $GM/c$, and $GM/c^3$, respectively. In this work, we follow a sign convention of the metric $(-, +, +, +)$, where the four velocities satisfy $u_\mu u^\mu = -1$. Throughout our study, Greek indices represent both space and time, i. e. $(0, 1, 2, 3)$; however, Roman indices represent only space, i. e. $(1, 2, 3)$.
\subsection{Initial conditions}
We set up a thin accretion disc threaded by the poloidal magnetic field. The disc extends in the radial direction from horizon up to $r_{\rm out}=500$ and in the polar direction from $\theta=0$ to $\theta=\pi$. The numerical domain for the reference run is resolved with an effective resolution of $2048\times 1024$. The initial disc density profile is calculated on the equatorial plane in spherical Boyer Lindquist (BL) coordinates following \cite{Novikov-Thorne1973, Page-Thorne1974}.

The thin disc approximation implies that all the matters occupy the equatorial plane of the coordinate system ($\theta=\pi/2$) and the $\theta$  component of the four-velocity $u^\mu$ is zero $(i.e., u^\theta = 0)$. Under this approximation, the accretion rate $(\dot{M_0})$, time average radiation flux $(F)$, and the time average torque $(W^r_\phi)$ are obtain as,
\begin{align}
\begin{aligned}
\dot{M}_0&=-2\pi r \Sigma u^r,\\
F&=\left(\dot{M}_0/4\pi r\right)f,\\
W^r_\phi&=\left(\dot{M}_0/2\pi r\right)\left[(E^\dagger-\Omega L^\dagger)/(-\Omega_{,r})\right]f,\\
\end{aligned}
\label{eq-02}
\end{align}
where $E^\dagger = -u_{t,e}$, $L^\dagger = u_{\phi,e}$, and $\Omega = u^{\phi}_e/u^{t}_e$. Quantities with subscript $e$ represent those defined on the equatorial plane. Considering the flow motion to be Keplerian $(u^r\sim0)$, the explicit expression of $f$ is obtain as
\begin{align}
f=-\Omega_{,r}(E^\dagger-\Omega L^\dagger)^{-2}\int_{r_{ms}}^r (E^\dagger-\Omega L^\dagger)L^\dagger_{,r}dr.
\label{eq-03}
\end{align}
Here, $r_{ms}$ is the marginally stable radius of the innermost stable circular orbit (ISCO): (radius at which $\frac{dE^\dagger}{dr}=\frac{dL^\dagger}{dr}=0$). The functional form of $f$ is given by

\begin{align}
\begin{aligned}
f =& \frac{3}{2} \frac{1}{ x^2 \left(2 a+x^3-3 x\right)}\bigg[ x - x_0 -\frac{3}{2}\ln\left(\frac{x}{x_0}\right) \\
&- \frac{3\left(s_1-a\right)^2}{s_1(s_1-s_2)(s_1-s_3)} \ln \left(\frac{x- s_1}{x_0-s_1}\right) \\
&- \frac{3\left(s_2-a\right)^2}{s_2(s_2-s_1)(s_2-s_3)}\ln \left(\frac{x- s_2}{x_0-s_2}\right) \\
&- \frac{3\left(s_3-a\right)^2}{s_3(s_3-s_1)(s_3-s_2)}\ln \left(\frac{x- s_3}{x_0-s_3}\right)\bigg], \\
\end{aligned}
\label{eq-04}
\end{align}
where $x=\sqrt{r}$ implying $x_0=r_{ms}^{1/2}$, and $s_1,s_2$ and $s_3$ are the roots of $s^3 - 3s + 2a=0$. The explicit form of $s_1,s_2$ and $s_3$ are given bellow,
\begin{align}
\begin{aligned}
s_1=&2 \cos \left(\frac{1}{3} \cos ^{-1}(a)-\frac{\pi }{3}\right),\\
s_2=&2 \cos \left(\frac{1}{3} \cos ^{-1}(a)+\frac{\pi }{3}\right),\\
s_3=&-2 \cos \left(\frac{1}{3} \cos ^{-1}(a)\right).\\
\end{aligned}
\label{eq-05}
\end{align} 
The temperature of the flow is obtained considering black-body distribution of radiation
\begin{align}
\frac{p_e}{\rho_e} \propto T_{\rm bb}(x)\propto F^{1/4} = \Theta_0 \left(\frac{f(x)}{x^2}\right)^{1/4},
\label{eq-06}
\end{align}
where $p_e$ and $\rho_e$ are the pressure and density at the equatorial plane. Considering polytropic equation of state $p = {\cal K}\rho^\Gamma$, the density of the fluid at the equatorial plane is obtain as
\begin{align}
\rho_e=\left(\frac{\Theta_0}{\cal K}\right)^{1/(\Gamma -1)} 
              \left(\frac{f(x)}{x^2}\right)^{1/(4(\Gamma - 1))},
\label{eq-07} 
\end{align}
where ${\cal K}$ is a constant related to the entropy of the flow. Eq. (\ref{eq-07}) gives the density profile at the equatorial plane ($\theta=\pi/2$). For off equatorial plane, we consider the density drops along the vertical direction following a normal distribution, therefore the general density profile can be written as,
\begin{align}
\rho(r,\theta) = \rho_e \exp\left(-\frac{\alpha^2 z^2}{H^2}\right); ~~ z=r\cos(\theta).
\label{eq-08}
\end{align}
To ensure the thin disc geometry, we choose $\alpha=2$, $H$ is the scale height of the accretion disc. 
We follow \citet{Riffert-Herold1995} and \citet{Peitz-Appl1997} to obtain the explicit expression of $H$,
\begin{align}
H^2 = \frac{p_e r^3}{\rho_e {\cal F}},
\label{eq-09}
\end{align}
where
$$
{\cal F}=\gamma_\phi^2\frac{\left(a^2+r^2\right)^2+2 a^2 \Delta }{\left(a^2+r^2\right)^2-2 a^2 \Delta },
$$
with $\gamma_\phi^2 = \left(1-\Omega\lambda\right)^{-1}$ and $\Delta = r^2-2 r + a^2$ where $\lambda = - u_{\phi, e}/u_{t,e}$.

To calculate $u^\phi$ at any point, we consider that the fluid element follow the geodesic equation given below :
\begin{align}
u^t u^t \Gamma^r_{tt}+ 2 u^t u^\phi \Gamma^r_{t\phi}+ u^\phi u^\phi \Gamma^r_{\phi\phi}=0,
\label{eq-10}
\end{align}
and the four velocities must satisfy
\begin{align}
g_{\mu\nu}u^\mu u^\nu = g_{\phi\phi}u^\phi u^\phi + g_{tt}u^t u^t + 2 g_{t\phi}u^t u^\phi = -1.
\label{eq-11}
\end{align}
Here, $g_{\mu\nu}$ is the metric tensor and $\Gamma^\alpha_{\beta \gamma}$ are the Christoffel symbols. Solving Eq. (\ref{eq-10}) and (\ref{eq-11}), we obtain the explicit form of $u^\phi$ and $u^t$, they are given by
\begin{align}
u^\phi(r,\theta) = \left(\frac{\cal A}{{\cal B}+ 2 {\cal C}^{1/2}}\right)^{1/2},
\label{eq-12}
\end{align}
where
$$
\begin{aligned}
{\cal A}=&\left(\Gamma^r_{tt}\right)^2,\\
{\cal B}=&g_{tt}\left(\Gamma^r_{tt}\Gamma^r_{\phi \phi}-2 {\Gamma^r_{t\phi}}^2\right)+2 g_{t\phi} \Gamma^r_{tt} \Gamma^r_{t\phi} - g_{\phi \phi } {\Gamma^r_{tt}}^2,\\
{\cal C}=&\left({\Gamma^r_{t\phi}}^2 - \Gamma^r_{tt} \Gamma^r_{\phi \phi}\right) (g_{t\phi} \Gamma^r_{tt}- g_{tt} \Gamma^r_{t\phi})^2.\\
\end{aligned}
$$
The expression of $u^\phi$ and $u^t$ on the equatorial plane ($\theta=\pi/2$),
\begin{align}
\begin{aligned}
u^\phi_e =& \frac{1}{\sqrt{x^3 \left(2 a+x^3-3 x\right)}},\\
u^t_e= & \frac{a+x^3}{\sqrt{x^3 \left(2 a+x^3-3 x\right)}}.\\
\end{aligned}
\label{eq-13}
\end{align}
Thus,  $\rho(r,\theta)$ and $u^\phi(r,\theta)$ obtain in Eq. (\ref{eq-08}) and Eq. (\ref{eq-12}) serve as initial conditions for the thin disc setup in our simulation models. Note that the code takes inputs in MKS (Modified Kerr-Schild) coordinates. Therefore, the initial conditions calculated at BL coordinate are properly transformed to MKS coordinate before supplying to the simulation models.

To ensure that the code can handle low density environments particularly close to the black-hole, a model for floor density and pressure is set as $\rho_{\rm flr}=\rho_{\rm min}r^{-1/2}$ and $p_{\rm flr}=p_{\rm min}r^{-3/2}$, with $\rho_{\rm min}=10^{-5}$ and $p_{\rm min}=10^{-7}$.  Additionally, we also set the maximum bound to the Lorentz factor to be $\gamma_{\rm max}=20$.
\subsection{Boundary conditions}
In our simulation models, we consider no-inflow boundary conditions at the radial boundaries. In contrast, the scalar variables and the radial vector components are symmetric at the boundaries with the polar axis. At the same time, the azimuthal and polar vector components are considered to be antisymmetric at the boundaries with the polar axis. Thus, our accretion disc is devoid of inflow at the outer edge of the accretion disc. Since our interest is to study the launching of jet and wind and the inner disc properties, therefore we run our simulations only up to $\sim 10 - 15\%$ of the outer edge rotation time, which corresponds to $\sim 400 - 450$ inner disc orbits.
\subsection{Parametric Models}
The initial poloidal magnetic field lines threading the accretion disc are prescribed using the vector potential ${\cal A}_\phi$. We follow \cite{Zanni-etal2007} to set the inclined field profile for the poloidal field lines. The explicit form of the vector potential is given by
\begin{align}
{\cal A}_\phi \propto \left(r \sin \theta\right)^{3/4} \frac{m^{5/4}}{\left(m^2 + \tan^{-2}(\theta-\pi/2)\right)^{5/8}},
\label{eq-14}
\end{align}
where the parameter $m$ determines the initial inclination of the poloidal field lines and the magnetic flux. It is noteworthy that the parameter $m$ plays a very crucial role in the launching of magneto-centrifugal disc winds \citep{Blandford-Payne1982}. The inclination angle is measured with respect to the equatorial plane, where a higher value of $m$ corresponds to a higher inclination angle. 

The magnetic field strength is determined by the choice of the plasma-$\beta$ parameter $\beta_{\rm inp} = p_{\rm gas}^{\rm max}/p_{\rm mag}^{\rm max}$. Here, the $p_{\rm gas}^{\rm max}$ and $p_{\rm mag}^{\rm max}$ are the maximum values of the gas pressure and the magnetic pressure, respectively, in the simulation domain. Additionally, we also compute the maximum plasma-$\beta$ parameter in the initial setup (i.e., $\beta_{\rm max}$). We obtain $\beta_{\rm max}$ on the equatorial plane at the radius of density maximum (i.e., at $r=r_{\rm max}$). Here the flow also has its maximum gas pressure $(p_{\rm gas}^{\rm max})$, but not necessarily the maximum value of magnetic pressure. We consider six simulation models with different values of $\beta_{\rm inp}$, respectively $\beta_{\rm max}$, and $m$.

We study the effect of input plasma-$\beta$ parameter and the inclination angle of the magnetic field lines on jet launching and disc-wind driving from an accretion disc around Kerr black hole with Kerr parameter $a=0.9375$. 
For this Kerr parameter, the radius of the ISCO and the maximum density radius are obtained as $r_{ms}=2.0442$, and $r_{\rm max}=2.97973$, respectively.

We consider model C with $m=0.1$ and $\beta_{\rm inp}=0.01$ as the reference run for the present study. Additionally, for comparison, we also perform the reference run with a higher effective resolution of $4096\times 2048$ (model C$_{\rm H}$). The details of various runs considered in our present work are given in table \ref{tab-01}.
\begin{table}
\centering
  \begin{tabular}{| l | c |c | c | c |}
    \hline
    Model & Effective resolution & $\beta_{\rm inp}$ & $\beta_{\rm max}$ & $m$ \\ 
    \hline
    A &  $2048\times 1024$ & 1 & 18844 & 0.1\\
    B &  $2048\times 1024$ &  0.1 &1884 & 0.1\\
    C &  $2048\times 1024$ & 0.01 &188 & 0.1\\
    C$_{\rm H}$ &  $4096\times 2048$ & 0.01 &  188 & 0.1\\
    D &  $2048\times 1024$ & 0.01 & 6163  & 0.4\\
    E &    $2048\times 1024$ & 0.01 & 13939  & 0.6\\
    F &  $2048\times 1024$ & 0.01 & 22657  & 0.8\\
    \hline
  \end{tabular}
\caption{The explicit values of effective resolution, input plasma-$\beta$, $m$, and $\beta_{\rm max}$ for different simulation models.}
\label{tab-01}
\end{table}

With the increase of $m$, the maximum value of the plasma-$\beta$ parameter ($\beta_{\rm max}$) increases. This essentially implies that, as the field lines become more vertical, the magnetic flux along the equatorial plane decreases, and thereby, the flow becomes less magnetized.

The density and pressure in the equatorial plane are initialised applying  $\Theta_0 = 0.001$ and ${\cal K}=0.1$ and we consider $\Gamma = 4/3$. As a consequence, we find that the maximum value for the aspect ratio obtained $\left(H/r\right)_{\rm max}\sim0.07$ at the outer boundary of the initial simulation setup ($r=r_{\rm out}$). 

\begin{figure}
\includegraphics[scale=0.3]{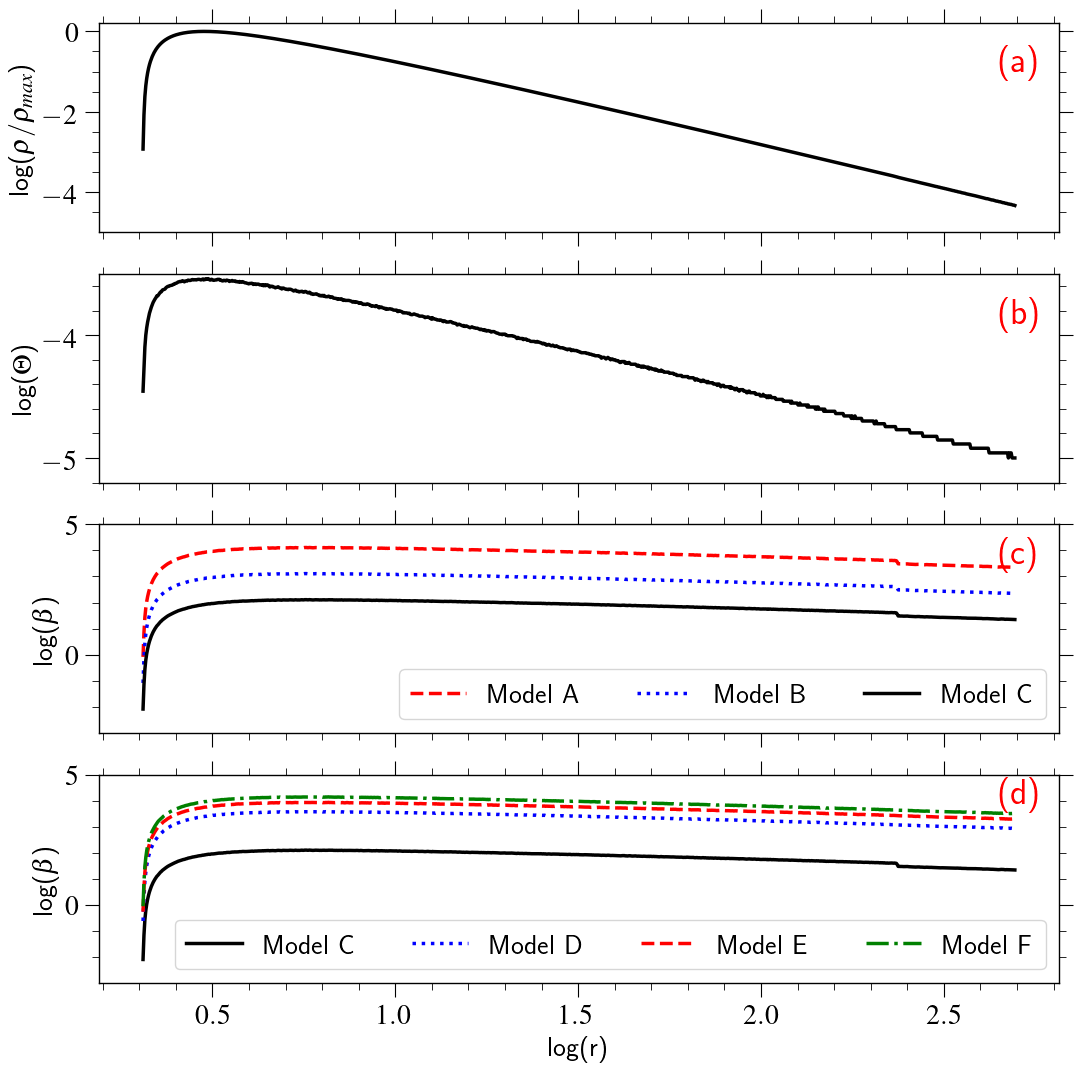}
\caption{Shown is (a) the  logarithmic normalize density profile $(\rho/\rho_{\rm max})$,  (b) logarithmic temperature ($\Theta = p/\rho$), and (c,d) logarithmic plasma-$\beta$ profile for different simulation models on the equatorial plane at time $t=0$.  See the text for details.}
\label{fig-initial}
\end{figure}
\begin{figure}
\centering
\includegraphics[scale=0.3]{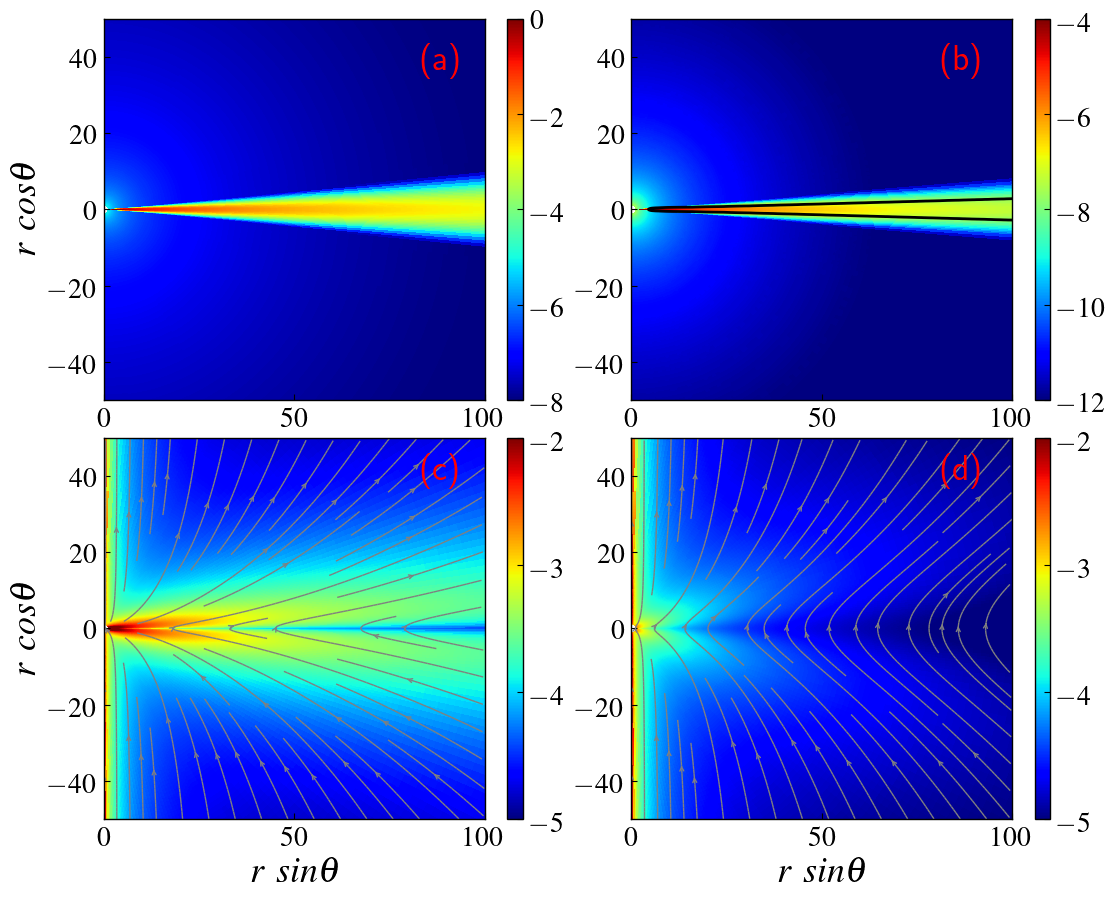}
\caption{Logarithmic normalized density $(\rho/\rho_{\rm max})$ (a) and 
         logarithmic gas pressure $(p_{\rm gas})$ (b) 
on the poloidal plane for the reference model at $t=0$. 
Solid black lines corresponds to the contour of plasma-$\beta=1$. 
The poloidal component of the magnetic field 
$B_{\rm p}=\sqrt{B^rB_r + B^\theta B_\theta}$ (in log-scale) and 
the poloidal field lines (grey lines) are shown for (c) model C $(m=0.1)$ and (d) model D $(m=0.4)$. See the text for details.}
\label{fig-initial2d}
\end{figure}
The initial radial profile for density $(\rho/\rho_{\rm max})$ (Eq. \ref{eq-08}) and temperature ($\Theta = p/\rho$) along the equatorial plane are shown in Fig.~\ref{fig-initial}a and Fig.~\ref{fig-initial}b, where $\rho_{\rm max}$ is the density calculated at the $r_{\rm max}$. The initial variation of plasma-$\beta$ along the radius on the equatorial plane for models A (black solid), B (blue dotted), and C (red dashed) are shown in \ref{fig-initial}c. 

The comparison for the initial radial profile for plasma-$\beta$ with different inclination parameters ($m$) along the equatorial plane is shown in Fig.~\ref{fig-initial}d.  Black solid, blue dotted, red dashed, and green dashed dot lines corresponds to Model C ($m=0.1$), D ($m=0.4$), E ($m=0.6$), and F ($m=0.8)$, respectively. As we mentioned earlier, it is clear from these profiles that with the increase of the inclination parameter, the equatorial disc plane becomes less magnetized. The role of different inclination parameters $(m)$ in governing the flow dynamics are also explored in this study.

In Fig. \ref{fig-initial2d}a and Fig. \ref{fig-initial2d}b, we show the initial density and the gas pressure distribution in the poloidal plane for our reference model, respectively. The solid black line in Fig. \ref{fig-initial2d}b corresponds to the contour with plasma-$\beta=1$. Note that the initial density and the gas pressure profiles are the same for all the models. The density profile near the equatorial plane follows Eq. (\ref{eq-08}). Far from the equatorial plane, the setup is filled by the floor density as described in the previous section (section 2.2). Most of the flow in the accretion disc is gas pressure-dominated. The flow near the black hole is magnetic pressure-dominated. As the gas pressure in the initial setup drops exponentially along the vertical direction, the flow far from the equatorial plane is magnetic pressure dominated.
 
In Fig. \ref{fig-initial2d}c and Fig. \ref{fig-initial2d}d, we compare the field structure and the poloidal component of the magnetic field ($B_{\rm p}=\sqrt{B^rB_r + B^\theta B_\theta}$) for model C $(m=0.1)$ and D $(m=0.4)$, respectively. With the increase of the inclination parameter, not only the inclination of the poloidal field line increase but also the strength of the poloidal component around the equatorial plane increases significantly. This increase in the magnetic field component increases initial magnetic flux in the simulation models. Also, due to the highly inclined field lines, models with a lower value of $m$ contain higher magnetic tension in the poloidal field lines. These properties of the inclination parameter $(m)$ motivates us to devise models C, D, E, and F by varying $m$ parameters. 

\subsection{Non-ideal and radiative effects}
\begin{figure*}
\centering
\includegraphics[scale=0.42]{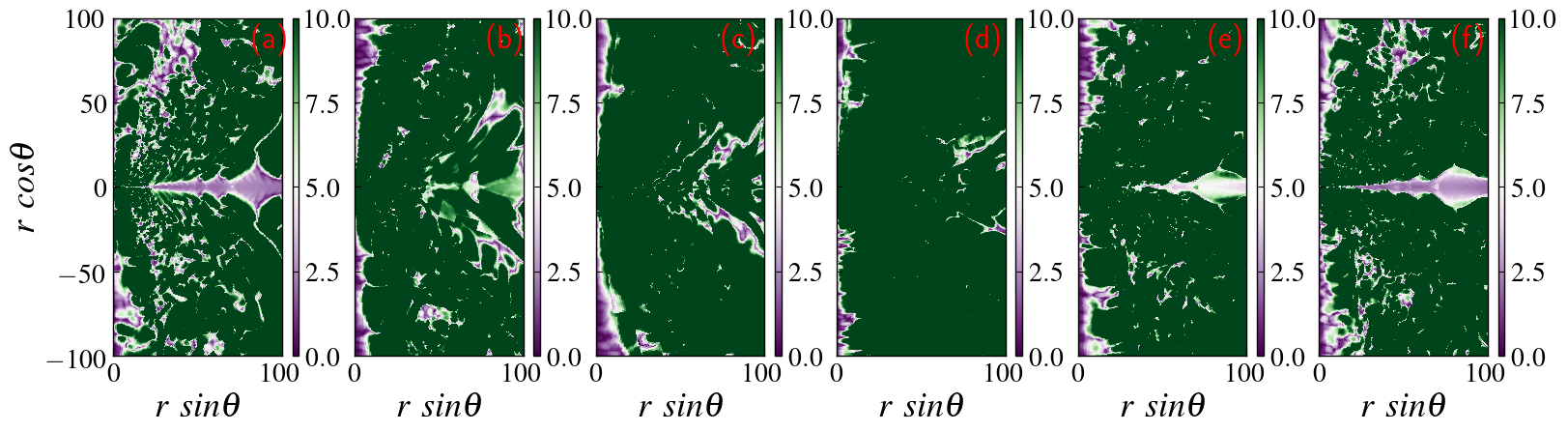}
\caption{MRI quality factor $Q_\theta$ in the poloidal plane for models
A, B, C, C$_{\rm H}$, D, and F 
at simulation time $t=4000$, 
in the panels (a), (b), (c), (d), (e), and (f), respectively.}
\label{fig-qtheta}
\end{figure*}
Non-ideal effects that contribute towards accretion processes include viscosity and resistivity. 
Viscous timescales govern the inflow of matter in the accretion disc. The standard approach of introducing viscosity in the accretion disc is given by \cite{Shakura-Sunyaev1973} using the $\alpha$ prescription. \cite{Balbus-Hawley1991, Balbus-Hawley1998} provided physical interpretation to the \cite{Shakura-Sunyaev1973} $\alpha$ prescription via the process of magneto-rotational instability (MRI). It was shown that MRI is one of the dominant and essential ingredients to transport angular momentum and in driving turbulence, specifically in magnetized accretion disc with differential rotation. In order to have physical accretion due to MRI, it is necessary to resolve the fastest growing mode along the vertical direction of the disc.
To ascertain that MRI is resolved for a chosen set of numerical resolutions, we define quality factor in terms of the wavelength $\lambda_\theta$ of the fastest growing MRI mode in $\theta$ direction as $Q_\theta = \lambda_\theta/\Delta x_\theta$ (see \cite{Takahashi2008,Siegel-etal2013,Porth-etal2019,Nathanail-etal2020}, for details). Here, $\lambda_\theta$ is given by
\begin{align}
\lambda_\theta = \frac{2\pi}{\sqrt{(\rho h + b^2)\Omega}}b^\mu e_\mu^{(\theta)},
\end{align}
and the grid resolution $\Delta x_\theta = \Delta x^\mu e_\mu^{(\theta)}$. These quantities are calculated at tetrad basis of the fluid frame $e_\mu^{\hat{(\alpha)}}$. Typically, this MRI mode will be resolved with values of $Q_\theta \gtrsim 6$ (see \cite{Sano-etal2004}). The figure (\ref{fig-qtheta}) shows the distribution of the quality factor $Q_\theta$ at time $t=4000$ within $r<100$ for different simulation models. 

With the increase of the strength of the magnetic field (decreasing $\beta_{\rm inp}$, decreasing $m$), the $\lambda_\theta$ increases. Accordingly, the distribution of $Q_\theta$ in the figure demonstrates that the fastest growing MRI mode is well resolved in models B, C, C$_{\rm H}$, and D ($Q_\theta \gtrsim 6$). However, for models A and D, the fastest growing MRI mode is well resolved only in the inner-most part of the disc. The outer part of the thin-disc is under-resolved for these two models.

Even though the simulations presented in this work do not have any explicit viscosity, the resolution particularly, in the inner-most region, allows for MRI-driven turbulence to help transport the angular momentum (e.g., the disc-wind). Additionally, the wavelength of the fastest growing MRI mode ($\lambda_\theta$) increases with the increase of the magnetic field strength. Therefore, this wavelength is better resolved for models with higher magnetic field strength (model C, C$_{\rm H}$). However, the growth rate of MRI-driven turbulence decreases with the strength of the magnetic field. In such a scenario, the efficiency of well resolved MRI-driven turbulence is reduced (see \cite{Bonanno-Urpin2008}), and other instabilities may drive the accretion process. For example, the magnetic-Rayleigh-Taylor instabilities (MRTI) is one of the possible candidates that drives turbulence in the magnetically dominated accretion disc (e.g., \cite{Igumenshchev2008, Avara-etal2016, Marshall-etal2018}, etc.) and may help in the transport angular momentum, particularly in model C.

The numerical viscosity present in the code also transports angular momentum, which may also lead to accretion only in the initial phase. However, after temporal evolution, the dominant mechanisms of transporting the angular momentum and accretion are different physical channels driven by turbulence and disc-winds. We will discuss the details of the accretion process in section 6.

Further, we also do not consider any explicit resistivity in the governing equations. In principle, that would prevent us from detecting any reconnection events. However, there is a resistivity present in our code due to the lack of infinite resolution in the simulation. Such numerical resistivity imitates the Ohmic resistivity of real plasma \citep{Kadowaki-etal2018}. It is to be noted that the resistivity of the models depends on the supplied resolution of the simulation. With such an ideal GRMHD code, it is intriguing to speculate the potential sites of magnetic reconnection, the formation of plasmoids, and its dynamical effects \citep{Nathanail-etal2020}. At the same time, the results observed due to the magnetic reconnections are phenomenologically consistent but essentially a numerical artifact \citep{Kadowaki-etal2018, Nathanail-etal2020}. Thus, we do not study accretion by varying the physical resistivity of the flow.  In order to capture the proper timescale and energetics of reconnections, one must include physical resistivity in the governing equations \citep{Qian-etal2017, Vourellis-etal2019, Ripperda-etal2020, Nathanail-etal2020}.  Nevertheless, there are ambiguities in resolving the energetics of reconnections, even in resistive magnetohydrodynamics \citep[and references therein]{Zweibel-Yamada2016}.

In the present study, the radiative effects are included only while prescribing the initial thin disc conditions using time-averaged radiative flux term (see Eq.~\ref{eq-02}) and not accounted for during the evolution. 
Including explicit radiation effects is interesting from the thermodynamics point of view, particularly quantifying disc luminosity and its impact on emission signatures. For example,  few GRMHD simulations have included an ad-hoc optical thin cooling to quantify the radiative efficiency of the thin magnetized disc and subsequent deviation from the NT model \citep{Noble-etal2009, Penna-etal2010, Avara-etal2016}.  
On the other hand, the present work primarily focuses on understanding the launching and dynamics of jets and winds from an underlying thin accretion disc. Such dynamical behaviors are governed by the forces that act on the fluid under consideration. Including explicit thermal cooling will alter the temperature and pressure, but its effect on driving jets and winds will be relatively less as compared to the dominant magnetic forces. In order to quantify the effect on disc thickness in the absence of explicit cooling and heating, we have evolved our initial conditions without magnetic fields for time $t = 10000$. We find that the initial disc structure does not change significantly (see Appendix A), implying that the prescribed initial condition without any inclusion of heating or cooling terms is well suited to study the impact of magnetic field on jet and wind driving. 

Further, we also have investigated the effect of varying disc height on the qualitative nature of disc and jet driving. 
The steady state thermodynamic properties governed by three parameters viz.,  $\Gamma$, ${\cal K}$, and $\Theta_0$ controls the initial disc's aspect ratio (i.e., variation of scale height $H$ with radius $r$). In particular, for parameters corresponding to model C (see Table~\ref{tab-01}), we have varied $\Theta_0$ from $10^{-4}$ to $5\times10^{-3}$ to obtain a variation in $(H/r)_{\rm max}$ from 0.02 to 0.16. 
The results obtained and subsequent discussion on these additional runs and its comparison with model C (our reference case) is presented in Appendix B.

\section{Dynamical Evolution}

\begin{figure*}
\includegraphics[scale=0.4]{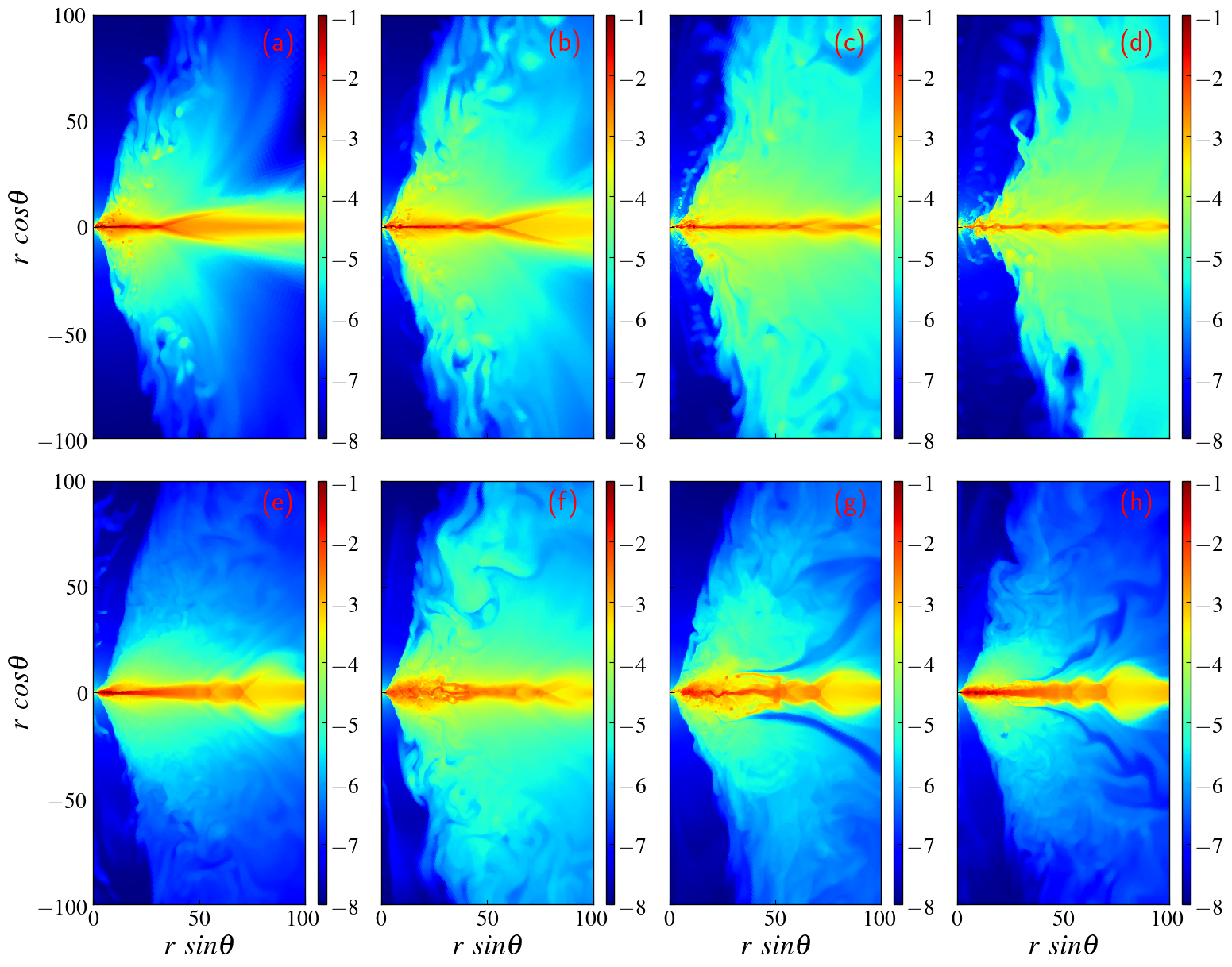}
\caption{Evolution of the logarithmic normalize density distribution
$(\rho/\rho_{\rm max})$ in the poloidal plane, 
for model C at $t=500$ (a), model C at $t=1000$ (b), model C at $t=2000$ (c),
model C at $t=4000$ (d), model A at $t=4000$ (e), model B at $t=4000$ (f), 
 model D at $t=4000$ (g), and model-F at $t=4000$ (h).}
\label{fig-density}
\end{figure*}

The temporal evolution of the normalised density profile for the reference run in logarithmic scale at four different simulation time $t=500, 1000, 2000$, and $4000$ is shown in the upper panels of Fig. \ref{fig-density}. 
The flow density structure in the simulation domain can be divided into three components: 
\begin{itemize}
    \item Low density funnel region near the rotation axis of the black hole.
     \item Disc-wind part occupying the off-equatorial part of the disc, which has a comparatively lower density than that of the accretion disc part.
    \item High-density thin accretion disc occupying the region near to the equatorial plane.
\end{itemize}

At the onset of the simulation runs, the disc rotational velocity winds the poloidal component of the magnetic field, resulting in the generation of the toroidal component ($B_{\rm tor}$). Subsequently, these magnetic fields play an integral role in transporting angular momentum (see section 6 for details). This eventually sets in the accretion of matter from the disc onto the central black hole. The matter accretion happens in an inside-out manner. As a  result, we observe a gradual depletion of high equatorial plane density with time and subsequent concentration towards the $\theta= \pi/2$ plane.

The logarithmic values of normalized density in the $r-z$ plane for models A, B, D, and F at time $t=4000$ are shown from left to right in the lower panels of Fig. \ref{fig-density}. The effect of decreasing input value of plasma-$\beta$ can be seen by comparing the flow structure for panels (e), (f), and (d) of Fig. \ref{fig-density}. In particular, we observe that with the decrease of the input plasma-$\beta$ parameter (model A $\rightarrow$ C) or with the increase of the initial magnetic field strength, the matter in the disc-wind region increases. 

Comparing the density structure for the reference run (in panel (d, $m=0.1$)) with those in panels (g, $m=0.4$) and (h, $m=0.8$) of Fig.~\ref{fig-density} shows the effects of the initial magnetic field structure at time $t = 4000$. We observe that the density in the disc wind decreases as the initial poloidal magnetic field structure becomes more vertical with increasing value of the inclination parameter $m$. We have discussed earlier that the initial magnetic field strength along the equatorial plane increases for more inclined field structures (thus also the magnetic flux that is carried by the disc). Thus, our results obtained for the models with varying inclination parameter $m$ are consistent with the results observed for a variation of the initial magnetic field strength.

We also observe a highly collimated and very low density flow material in the funnel region (seen as dark blue color in Fig. \ref{fig-density}) close to the black hole. The temporal evolution demonstrates that the low density flow widens but retains the parabolic structure throughout the simulation time. We fit the boundary of the funnel region with a power-law $(Z_j \propto X_j^\xi)$, where $Z_j = r_j~\cos\theta_j$ and $X_j = r_j~\sin\theta_j$, the suffix $j$ indicates that the quantities are calculated at the boundary of the funnel region. The best fit power-law index $\xi$ obtain for the reference model are  $\xi=1.7, 1.8, 2.3,$ and $2.6$ at simulation times $t=500, 1000, 2000,$ and $4000$, respectively. The funnel region is surrounded by the disc-wind outflow, which has a relatively higher density than that of the funnel region. As time evolves, disc-wind outflow material fills up and is seen to be covering the whole of the numerical domain. 

In order to understand the structure of the funnel region, we study the evolution of the power-law index $\xi$. We observe that it reaches a quasi-steady value. For example, in the case of model B, the power-law index remains within a range of $\xi \sim 1.8-2.0$ for a sufficiently long time $t\sim 300-7000$.  Similar quasi-steady behavior of funnel region is also seen for other models (A, D, E, and F), and the power-law index remains within the same range, i.e., $\xi\sim 1.8-2.0$. 
However, for model C the funnel region behaves in a rather chaotic manner, and the quasi-steady state of the funnel region is for a very short period of time when $\xi$ attains values in the range $\sim 1.8-2$. This implies that the funnel region during the quasi-steady state attains a similar shape irrespective of the choice of initial inclination angle and initial magnetic field strength. Such a shape is less collimated than that of a genuine parabolic shape $(Z_j \propto X_j^2)$ in accordance with \cite{Nakamura-etal2018}. 
It should be noted that the initial strength and the inclination of the field line play crucial roles in setting up the turbulent disc-wind region (see section 5), which surrounds the inner funnel region. The strength of such disc-wind controls the temporal extent of quasi-steady behavior of inner BZ-jet. In particular,  with stronger and inclined filed, the jet structure is more turbulent, while the jet structure is more stable with weaker and vertical filed. We also note that the funnel structure of the BZ-jet with a very weak and vertical field is not well developed (see section 4).

In summary, both parameters that determine the magnetic field structure are crucial in understanding the flow dynamics. 
A detailed description of the effects of these parameters on other dynamical quantities pertaining to launching of BZ jet, driving of BP winds and accretion flow in thin disc is described in details in the following sections.   

\section{Launching of Blandford \& Znajek-jet}

\begin{figure*}
\includegraphics[scale=0.4]{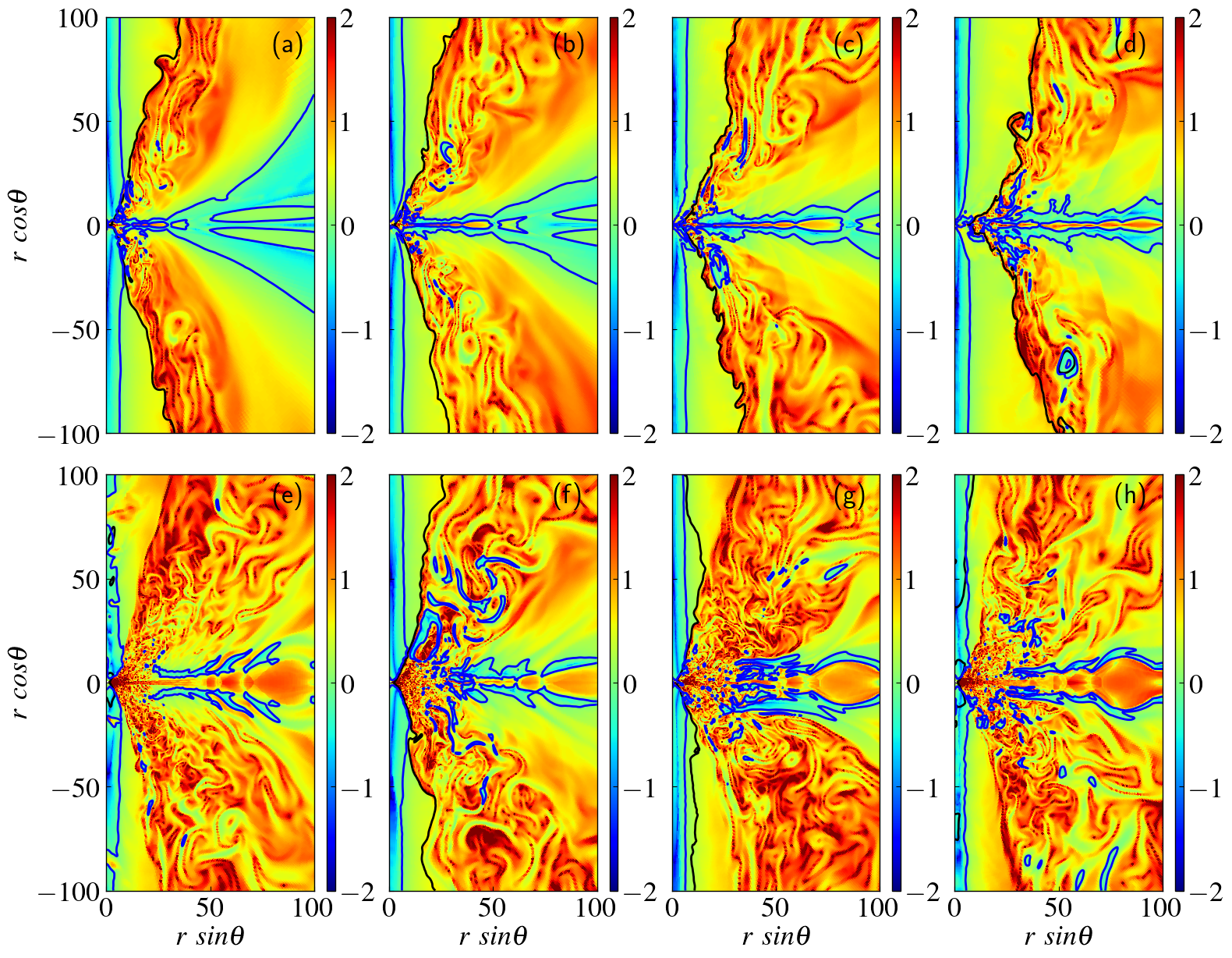}
\caption{Evolution of the logarithmic poloidal Alfv\'enic Mach number 
$(M_{\rm A, p})$ distribution in the poloidal plane for
model-C at $t=500$ (a), 
model C at $t=1000$ (b), 
model C at $t=2000$ (c), 
model C at $t=4000$ (d), 
model A at $t=4000$ (e), 
model B at $t=4000$ (f), 
model D at $t=4000$ (g), and 
model F at $t=4000$ (h). 
The solid black and blue lines correspond to contours of 
$\sigma=1$ and $M_{\rm A, p}=1$.}
\label{fig-machv}
\end{figure*}
\begin{figure*}
\includegraphics[scale=1.1]{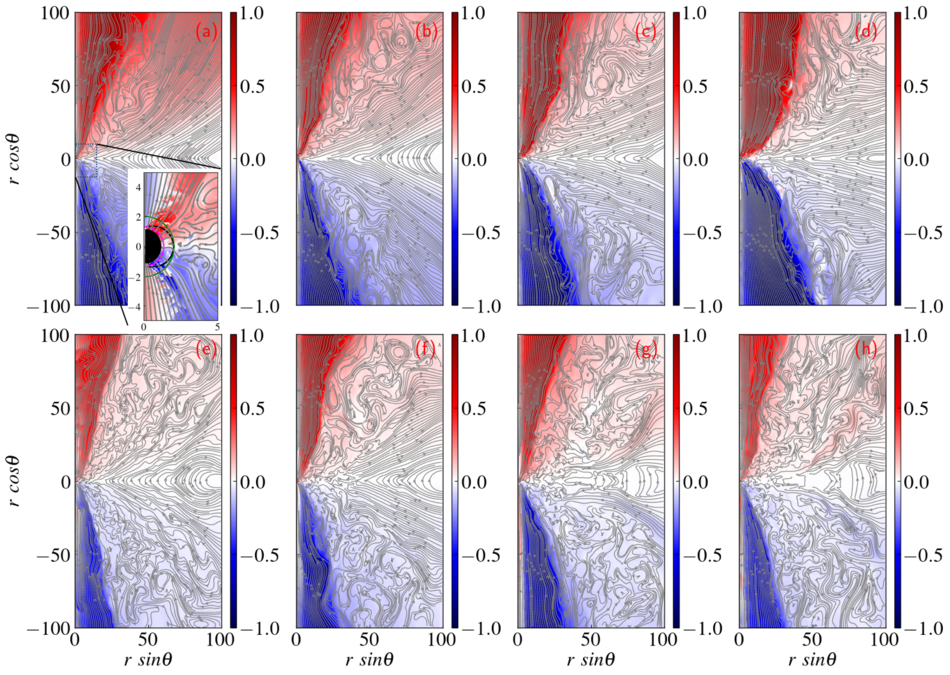}
\caption{Evolution of the vertical velocity profile $(v^z)$ in the poloidal plane
for model C at $t=500$ (a), 
model C at $t=1000$ (b), 
model C at $t=2000$ (c), 
model C at $t=4000$ (d), 
model A at $t=4000$ (e), 
model B at $t=4000$ (f), 
model D at $t=4000$ (g), and 
model F at $t=4000$ (h). 
The grey lines correspond to the polar field lines. In the onset of subfigure (a), we display a zoomed version of the same, with the magenta semicircle marking the horizon, the black line marking the ergosphere, and the radius of ISCO is indicated by the green line.}
\label{fig-fieldlines}
\end{figure*}
\begin{figure*}
\includegraphics[scale=0.4]{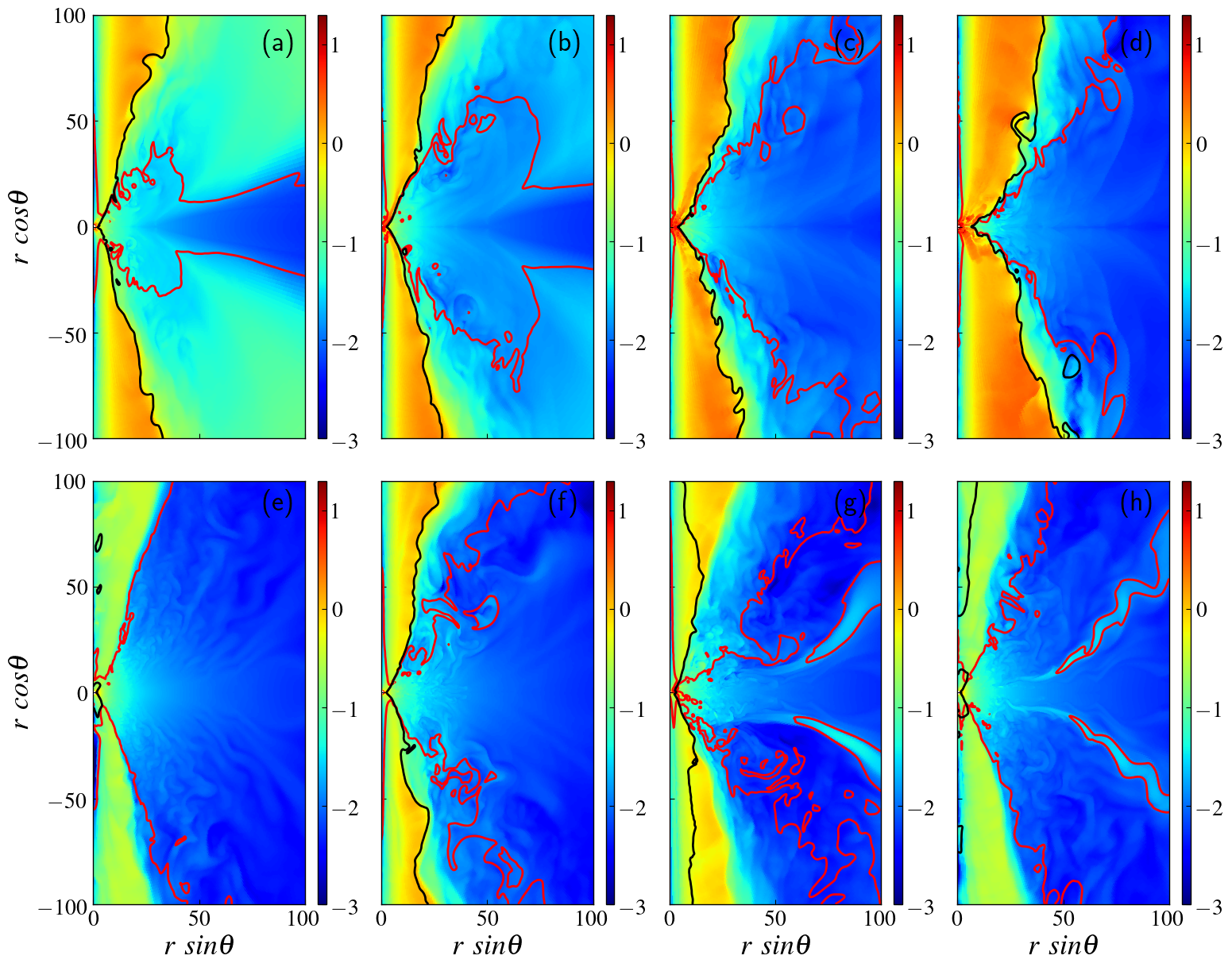}
\caption{Evolution of the logarithmic $\gamma-1$ distribution in the poloidal plane, (a) model C at $t=500$, (b) model C at $t=1000$, (c) model C at $t=2000$, (d) model C at $t=4000$, (e) model A at $t=4000$, (f) model B at $t=4000$, (g) model D at $t=4000$, and (h) model F at $t=4000$. The solid black and red line corresponds to $\sigma=1$ and ${\cal E}=1$ contour, respectively.}
\label{fig-lfac}
\end{figure*}

To study the launching mechanism of the Blandford \& Znajek-jet (BZ-jet) from the ergosphere of the black hole, we show the temporal evolution of the poloidal Alfv\'enic Mach number ($M_{\rm A, p}$) distribution for the reference run in logarithmic scale at four different simulation times $t=500, 1000, 2000$, and $4000$ in the upper panels of Fig. \ref{fig-machv}, where solid black and blue lines depict the contours for magnetisation $\sigma=b^2/\rho=1$ and $M_{\rm A, p}=1$, respectively. We calculate the Alfv\'enic velocity by $c^2_{\rm a}=B_p^2/(\rho h + B_p^2)$, and the poloidal Alfv\'enic Mach number is obtained as $M_{\rm A, p}=u_p/c_a$, where $u_p^2 = u^ru_r + u^\theta u_\theta$ and $B_p^2 = B^rB_r + B^\theta B_\theta$ \citep{Qian-etal2018}. The temporal evolution of the vertical velocity ($v^z$) along with the poloidal magnetic field lines is shown in Fig. \ref{fig-fieldlines}. We also show the temporal evolution of Lorentz factor $\gamma$ (Fig. \ref{fig-lfac}) where the solid black and  red line correspond to the contour of magnetisation $\sigma=1$ and Bernoulli parameter ${\cal E}(=-hu_t)=1$ with $h$ being the specific enthalpy of the fluid.

In the reference model, at time $t=0$, the accretion flow is mostly sub-Alfv\'enic, except in the high-density equatorial plane region. At time $t=500$, we observe that the funnel region is occupied by super-Alfv\'enic flow ($M_{\rm A, p}>1$) (see the region between blue and black lines around the axis of rotation in Fig. \ref{fig-machv}a). In the reference model with a higher value of initial magnetic field strength, we observe that the super-Aflv\'enic flow attains a Lorentz factor up to $\gamma \sim 10$\footnote{The Lorentz factor and the maximum magnetisation of the jet within the funnel region depends on the choice of the floor density model. We have adopted the default density floor model as described in section 2.2 and kept it the same for all the runs. \label{footnote-01}} (see the dark reddish region of Fig. \ref{fig-lfac}a - Fig. \ref{fig-lfac}d) far away from the black hole where magnetisation $\sigma>1$, and Bernoulli parameter ${\cal E}>1$. In comparison, flow close to the black hole is sub-Alfv\'enic ($M_{\rm A, p}<1$) and gravitationally bound ${\cal E}<1$. As an inset of Fig. \ref{fig-fieldlines}a, we have drawn three semicircles in magenta, black, and green lines corresponding to the event horizon, ergosphere, and the radius of ISCO, respectively. In the figure, we observe that the field lines are polar and rooted in the ergo-sphere of the black hole. The above dynamical features affirm that the BZ-jet is launched from the region close to the black hole \citep{Blandford-Znajek1977, Komissarov-Barkov2009}. 

The BZ-jet broadens, and its shape becomes parabolic with the temporal evolution as evident from the upper panels of Fig~\ref{fig-machv} (follow the $\sigma=1$ contour (solid black line)). 
One also observes that a relatively higher density disc-wind surrounds the funnel region with velocities ranging from $0.1-0.2c$ (see top panels of Fig.~\ref{fig-fieldlines}). 
In particular, disc-wind takes out the material around the jet base and reduces gas pressure support. 
The modification of the field lines due to the temporal evolution indicates the increase in the toroidal component of the magnetic field in the disc-wind region. 
Further, we also observe that the disc wind that shapes the funnel region has lower poloidal magnetic field strength (see section 7 for more discussion). 
The gain of the toroidal magnetic field in the disc-wind region results in a stronger tension force that could be responsible for the force balance that controls the shape of the BZ-jet.  

We further study the effect of the strength of the initial magnetic field on the BZ-jet by comparing the panels (d), (e), and (f) of Fig. \ref{fig-machv} - Fig. \ref{fig-lfac}. The panels (d), (e) and (f) represent models C, A, and B respectively at time $t = 4000$. It is evident that the area inside of $\sigma>1$ increases with the increase of the strength of the magnetic field (model A $\rightarrow$ C). The maximum value of the magnetisation obtain in a simulation model also decreases with input plasma-$\beta$ parameter. For example, the maximum value of magnetisation at simulation time $t=4000$ for models A, B and C are $\sigma=7.80$, $\sigma=237.00$, and $\sigma=1007.00$ $^{\rm \ref{footnote-01}}$, respectively. The vertical velocity and the Lorentz factor within the funnel region decrease with the increase of initial plasma-$\beta$ parameter (see same panels of figures \ref{fig-fieldlines} and \ref{fig-lfac}). The maximum value of Lorentz factor observed in the funnel region for models A, B, and C are $\gamma \sim 1.4$, $\gamma \sim 2.5$, and $\gamma \sim 10.0$, respectively (see Fig. \ref{fig-lfac}d-f).
Such a variation of Lorentz factor $\gamma$ for the BZ jet is primarily due to the increase in the initial strength of the magnetic field with different models. Additionally, it should also be noted that the density in the relativistic BZ jet could depend on the model adopted for the floor density, particularly very close to the black hole.

The various flow properties for models with increasing value of the inclination parameter ($m$) of the initial magnetic field are compared at time $t = 4000$ in panels (d), (g) and (h) of Fig. \ref{fig-machv} - Fig. \ref{fig-lfac}. The lower value of parameter $m$, with the same input plasma-$\beta$ parameter signifies more inclined poloidal field lines at the beginning of the simulation (Fig. \ref{fig-initial2d}c,d). The models with a lower value of inclination parameter $(m)$ contain a more initial magnetic flux in the simulation domain, enabling a stronger BZ-process. The area with active BZ-process (area within magnetization $\sigma>1$ contour) decreases with an increase of $m$. The maximum value of the magnetization decreases by two orders of magnitude with increase in $m$ from $0.1$ to $0.8$.

The stronger magnetic field drives the BZ-jet efficiently. Consequently, we find that the maximum value of the vertical velocity in the BZ-jet decreases with the increase of $m$ (see in Fig. \ref{fig-fieldlines}). With the decrease of the vertical velocity in the BZ-jet, the Lorentz factor of the BZ-jet also decreases with the increase of $m$. The maximum value of the Lorentz factor at $t=4000$ for models C, D, and F are  $\gamma \sim 10.0$, $\gamma\sim 2.2$, and $\gamma \sim 1.5$, respectively  (see Fig. \ref{fig-lfac}d, g, and h).

In summary, the BZ-jet is characterized by the low density flow in the funnel region and is dominated by poloidal field lines that are anchored in the ergosphere of the rotating black hole. The increase of initial magnetic field strength results in the launching of highly relativistic BZ-jets with Lorentz factor reaching up to $\gamma\sim 10$. With a fixed value of the input plasma-$\beta$ parameter, highly inclined field lines (i.e., a lower value of $m$) signifies a stronger poloidal magnetic field at the beginning of the simulation. Thus with the temporal evolution, the BZ-jet becomes less magnetized and relatively slower with initial vertical fields.

For the reference model that represents a highly magnetized and inclined field structure, the temporal evolution shows a modest change from the initial field configuration. Such a structure signifies a magnetically arrested disc (MAD) structure \citep{Narayan-etal2003, McKinney-etal2012}, and MAD is known to be an optimal configuration for BZ-mechanism \citep{McKinney-etal2012}. In our study, we also observe maximum BZ-jet activities in model C in comparison to other models (see section 6 for more detail).

\begin{figure*}
\includegraphics[scale=0.8]{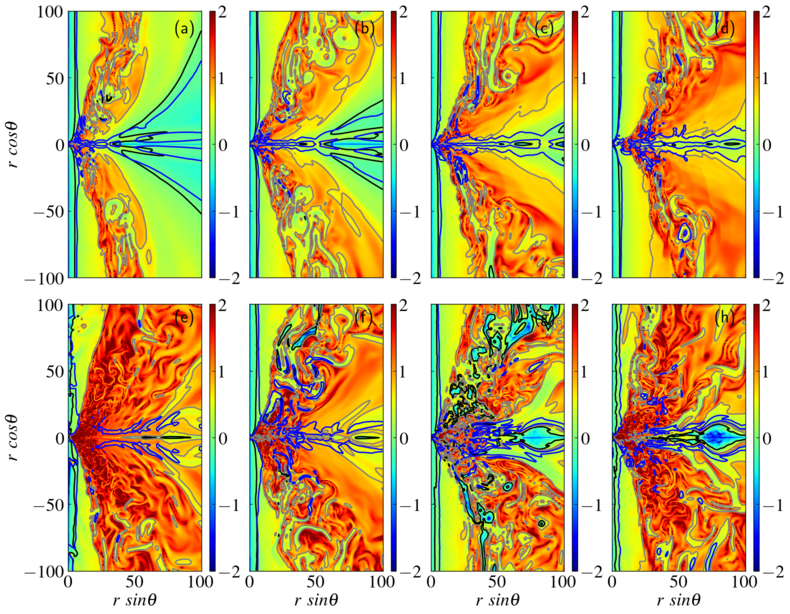}
\caption{Evolution of the logarithmic ratio of $B_{\rm tor}/B_{\rm p}$ 
distribution in the poloidal plane for 
model C at $t=500$ (a), 
model C at $t=1000$ (b), 
model C at $t=2000$ (c), 
model C at $t=4000$ (d), 
model A at $t=4000$ (e), 
model B at $t=4000$ (f), 
model D at $t=4000$ (g), and 
model F at $t=4000$ (h). 
The solid black, grey, and blue lines correspond to contours of 
$B_{\rm tor}/B_{\rm p}=1$,  $5$, and $M_{\rm A, p}=1$, respectively. }
\label{fig-btorbp}
\end{figure*}
\begin{figure}
\includegraphics[scale=0.43]{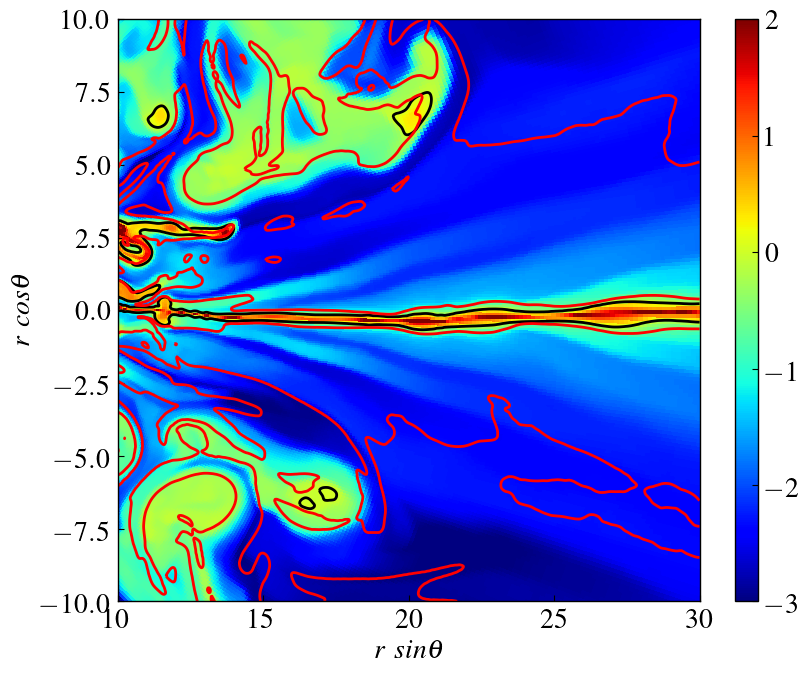}
\caption{Logarithmic plasma-$\beta$ distribution in the poloidal plane for
model C at time $t=2000$,
corresponding to $\simeq 2$ rotations of the disc at $r=30$. 
The black and red lines correspond to the contours $\beta=1$, and Alfv\'enic Mach number $M_{\rm A, p}=1$, respectively.}
\label{fig-bp-launc}
\end{figure}

\section{Driving disc-winds}
In our reference model, we find prominent features of disc-wind outflows. To further study their dynamics, we again consider Fig.~\ref{fig-machv} - Fig.~\ref{fig-lfac}. Along with these figures we also show the ratio of  $B_{\rm tor}/B_{\rm p}$ in Fig.~\ref{fig-btorbp} in the $r-z$ plane, following a convention of panels similar to the one adopted earlier figures. 
In Fig.~\ref{fig-btorbp}, the solid black, grey, and blue lines correspond to contours of $B_{\rm tor}/B_{\rm p}=1$,  $5$, and $M_{\rm A, p}=1$, respectively. The components of magnetic fields are estimated using $B^2_{\rm tor}=B^\phi B_\phi$, and $B^2_p=B^r B_r + B_\theta B^\theta$. 

Figure~\ref{fig-machv} suggests that in the reference model for radius $r\gtrsim 10$, the disc-wind leaves the accretion disc surface with sub-Alfv\'enic velocity and becomes super-Alfv\'enic after traveling a certain distance along the z-direction. Inside the thin disc, the flow is super-Alfv\'enic $M_{\rm A, p}>1$ and the ratio $B_{\rm tor}/B_{\rm p}\lesssim 1$. In the region with sub-Alfv\'enic ($M_{\rm A, p}<1$) poloidal velocity,  the ratio $B_{\rm tor}/B_{\rm p}$ is of the order of unity ($\le 5$). We show logarithmic plasma-$\beta$ distribution in the poloidal plane for the model C at time $t=2000$ corresponding to $\sim 2$ rotations of the disc at $r=30$ (Fig.~\ref{fig-bp-launc}). The figure also shows contours of $\beta=1$ (black) and Alfv\'enic Mach number $M_{\rm A, p}=1$ (red). We observe that as the equatorial plane remains gas pressure dominated ($\beta>1$), away from the equatorial plane, the plasma-$\beta$ drops drastically, and flow becomes magnetically dominated ($\beta<1$). In the magnetically dominated region, we observe the flow to be sub-Alfv\'enic ($M_{\rm A, p}<1$).
This strongly suggests that the matter is magnetically driven in the form of wind. This process resembles with the properties of disc-wind as suggested by \cite{Blandford-Payne1982} (BP disc-wind). Further, for radius $r\lesssim 10$, we observe $B_{\rm tor}$ dominated disc-wind in panel (e)-(h) of Fig. \ref{fig-btorbp}. In $B_{\rm tor}$ dominated ($B_{\rm tor}/B_{\rm p}\ge 5$) disc-wind, the disc material leaves the underlying surface due to the gradient of toroidal magnetic pressure. 

Comparing the results from panel (d), (e), and (f) of Fig.  \ref{fig-machv} - Fig. \ref{fig-btorbp}, the impact of the magnetic field strength on the evolution of the disc-wind can be studied. 
In these figures, results are shown for models C, A, and B, respectively, at simulation time $t=4000$. A stronger magnetic field is better suited to guide the motion of the flow, for models with a lower value of $\beta_{\rm max}$, the simulation starts with a stronger poloidal magnetic field, and the flow tends to follow the poloidal field lines. Therefore, in the presence of the stronger poloidal field, the MRI is suppressed and can not amplify the toroidal component of the magnetic field efficiently \citep{Begelman-Pringle2007}. 

\begin{figure}
\includegraphics[scale=0.35]{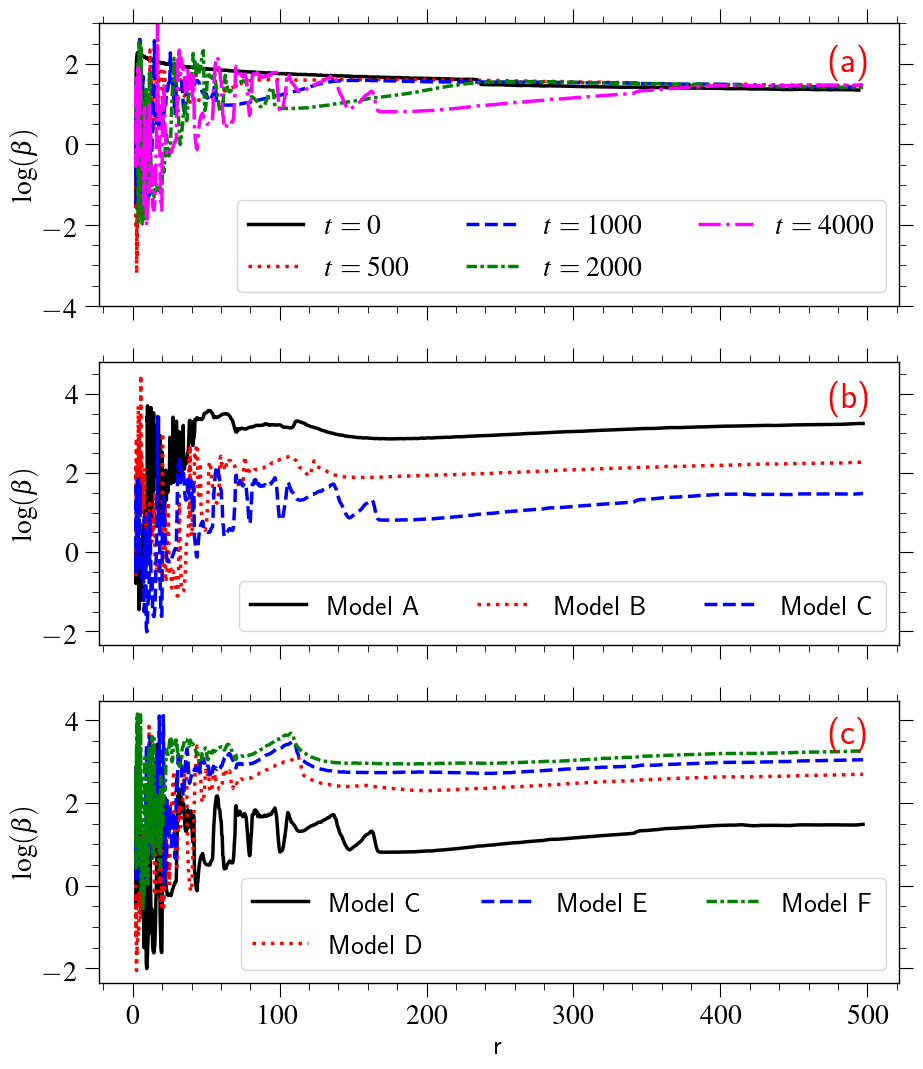}
\caption{Shown is the (a) plasma $\beta$ profile along the equatorial 
plane for different simulation times for the reference model $t=0$ (black solid), $500$ (red dotted), $1000$ (blue dashed), $2000$ (green dot-dashed) and $4000$ (magenta dot long dashed). 
Shown is also the (b) comparison between the plasma-$\beta$ profile along
the equatorial plane for Model A (black solid), B (red dotted), and C (blue dashed) at simulation time $t=4000$, 
and the (c) comparison between the plasma-$\beta$ profile along 
the equatorial plane for Model C (black solid), D (red dotted), E (blue dashed), and F (green dot-dashed) at simulation time $t=4000$. }
\label{fig-beta-evo}
\end{figure}

These facts reflect well in the $B_{\rm tor}/B_{\rm p}$ distribution of our simulation models. 
Indeed, we see the ratio of $B_{\rm tor}/B_{\rm p}$ decreasing with decreasing magnetic field strength (model C $\rightarrow$ A, see Fig.~\ref{fig-btorbp}d-f). Also we observe that with the increase of magnetic field strength, the area with active BP disc-wind increases, while the extent of $B_{\rm tor}$ dominated disc-wind decreases (model A $\rightarrow$ C). Further, the accretion flow becomes less turbulent with the lower value of $\beta_{\rm max}$ due to the strong initial poloidal magnetic field (see Fig. \ref{fig-fieldlines}). In order to understand it in detail, we show the evolution of plasma-$\beta$ profile on the equatorial plane for the reference model and comparison of the same for different models in panels Fig. \ref{fig-beta-evo}a-c. The disc equatorial plane or the thin disc is mostly gas pressure dominated ($\beta\gg1$).

With the temporal evolution, due to the generation of the toroidal component of the magnetic field, the value of $\beta$ decreases. The radial profile of $\beta$ at the equatorial plane shows signature of turbulence in the disc. We observe regions of higher magnetic pressure in between two higher gas pressure regions. They signify the turbulent amplification of the toroidal component of the magnetic fields. Such areas with a higher value of $B_{\rm tor}$ are responsible for driving turbulent disc-winds in these regions due to toroidal pressure gradients. 

As the rotation starts to progress towards the outer edge of the simulation domain, the plasma-$\beta$ profile also gets modified. However, the plasma-$\beta$ profile at the outer edge remains unchanged. In Fig.\ref{fig-beta-evo}c, we observe that the magnetic pressure along the equatorial plane is always higher for model C compared to the other models (Model A, B). Such a high magnetic pressure helps in launching the BP disc-wind in model C. In Model A and B, due to the weak magnetic pressure at the equatorial plane, MRI turbulence becomes efficient that leads to a broad $B_{\rm tor}$ dominated disc-wind as shown in Fig. \ref{fig-btorbp}d-f.

By comparing the results from panel (d), (g), and (h) of Fig.  \ref{fig-machv} - Fig.  \ref{fig-btorbp}, we study the effect of the structure of the initial structure of field lines on the disc-wind. In these figures, results are shown for models C $(m=0.1)$, D $(m=0.4)$, and F $(m=0.8)$, respectively at simulation time $t=4000$. 
A lower $m$ provides initial field lines that are more inclined towards the equatorial plane. There are two essential consequences. One is that a stronger curvature force directing radially outwards is acting on the disk material. Secondly, due to proximity in cases with higher inclination, the poloidal field lines can reconnect at the apex points.

\begin{figure}
\includegraphics[scale=0.4]{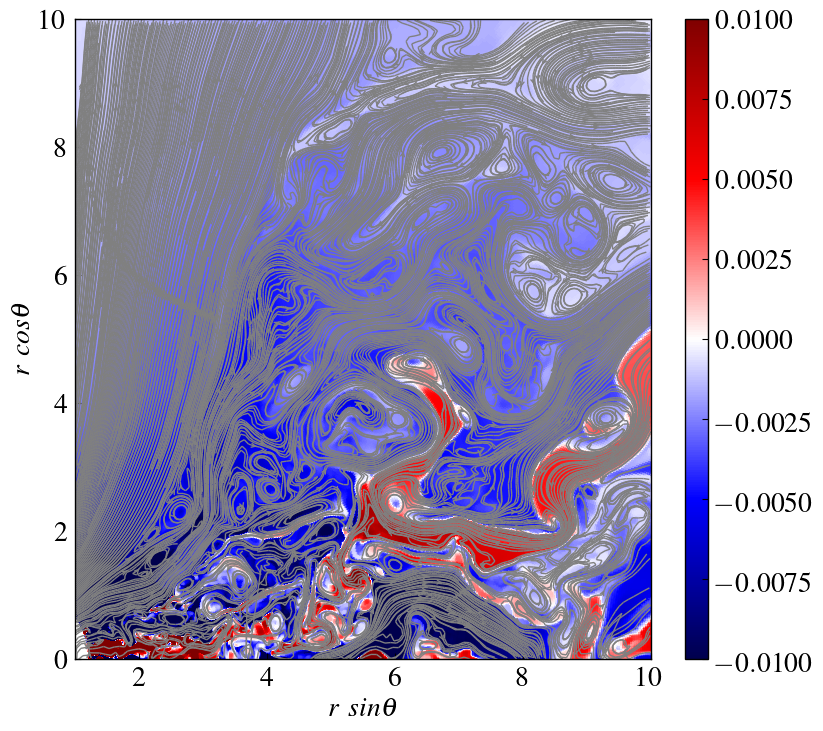}
\caption{Shown is the $B^\phi$ component of the magnetic fields and the poloidal field lines for model C$_{\rm H}$ at a simulation time $t=1000$. }
\label{fig-plasmoids}
\end{figure}

Around the apex point, the oppositely polar field lines are squeezed, leading to the reconnection events in the vicinity of the black hole. Also, due to turbulence, the amplified toroidal component $B_\phi$ changes its polarity driving reconnection (see Fig. \ref{fig-plasmoids})\citep{Nathanail-etal2020}. The distribution of toroidal component of the magnetic field along with the filed lines for mode C$_{\rm H}$ at simulation time $t=1000$ is shown in Fig.~\ref{fig-plasmoids}. We see that the presence of opposite polarity of the toroidal component of the magnetic field.

The growth of the reconnection depends on the resistivity present in the code that depends on the resolution. As a consequence of reconnection, we observe the formation of plasmoids and plasmoid chains due to tearing instability \citep[references therein]{Pucci-Velli2014,DelZanna-etal2016,Koide-etal2019}. These plasmoids are developed in the vicinity of the black hole. Subsequently, the plasmoids are advected in the form of current loops with the disc-wind as they are gravitationally unbound in these regions (${\cal E}>1$). These loops help to advect the toroidal component of the magnetic field with the disc-wind. 

With the increase of $m$, the angle of the apex point increases (i.e., field lines become more vertical and parallel to the Z-axis), and the strength of the poloidal component of the magnetic field near the equatorial plane also reduces (Fig. \ref{fig-initial2d}c, d). Therefore, in the simulation model with a higher value of $m$, the rotational motion of flow can guide the initial poloidal field lines efficiently to generate a stronger toroidal component of the magnetic field $B_{\rm tor}$. Also, the weaker magnetic field is suitable for MRI turbulence, which further amplifies the $B_{\rm tor}$ component. Therefore, we observe that the value of $B_{\rm tor}/B_{\rm p}$ obtained at simulation time $t=4000$, increases with the increase of $m$ (Fig. \ref{fig-btorbp}d, g, and h).

In Fig. \ref{fig-beta-evo}c, we compare the plasma-$\beta$ at the equatorial plane for models with different inclination parameters ($m$). With the increase of the $m$, the equatorial plane becomes less magnetized (higher $\beta$), which signifies less magnetic flux due to the poloidal component of the magnetic field, which plays the key role in magneto-centrifugal acceleration. 

A stronger poloidal magnetic field component near the disc equatorial plane results in stronger magnetization near the thin accretion disc. That helps in the launching of BP disc-wind through magneto-centrifugal acceleration. Consequently, we observe that with the increase in inclination parameter $m$, the area with active BP disc-wind decreases. Further, the toroidal component is generated due to presence of relatively more efficient MRI within the thin-disc. Therefore, the area with active $B_{\rm tor}$ dominated disc-wind increases with it. The presence of the large MRI active area for the models with higher inclination parameters ($m$) results in broadening of turbulent $B_{\rm tor}$ dominated disc-wind. As a result, we observe in increasing of turbulent nature of the field lines and area with $B_{\rm tor}/B_{\rm p}>1$  with the increase of $m$  (see Fig. \ref{fig-btorbp}d, g, and h).

In summary, the initial value of the input plasma-$\beta$ parameter and the inclination parameter of the initial field structure ($m$) both play crucial roles in deciding the nature of the disc-wind. 
With the increase of the strength and inclination (lower $m$) of the initial magnetic field, flow evolves from a large amount of initial magnetic flux along the accretion disc. 
These parameters decide the driving mechanism for the disc-wind. In models with higher values of $\beta_{\rm max}$ and $m$, the dominated disc-wind driving mechanism is toroidal magnetic pressure. While, in the models with lower $\beta_{\rm max}$ and $m$, the large scale magnetic stress plays a key role in driving the disc-winds. These winds are often referred to as magneto-centrifugally driven disc-wind.
\section{Disc-wind-jet connections}
In this section, we study the relationships between the disc, disc-wind, and jet. We first compare the mass flux rates of the different components. 
In general, the mass flux rate is written as follows,
\begin{align}
\dot{M}= 2\pi \int \sqrt{-g}\rho u^r d\theta.
\label{eq-15}
\end{align}
In order to calculate the mass flux in the BZ-jet $(\dot{M}_{\rm jet})$, we integrate Eq. (\ref{eq-15}) over the arc at a radius $r=50$ considering all locations with $\sigma>1$ or $\eta>2$, and ${\cal E}>1$, where  $\eta$ is the efficiency factor of the Poynting flux, and is defined  as, $\eta = -(T^r_t - \rho u^r)/(\rho u^r)$  \citep{Nathanail-etal2020}. Further, it is to be noted that the mass flux in the BZ-jet $(\dot{M}_{\rm jet})$ will also have contribution due to the prescribed floor density model.
We calculate disc-wind mass loss rate $(\dot{M}_{\rm wind})$ at the same radius ($r=50$), obtained by integrating Eq. (\ref{eq-15}) over the arc where $\sigma<1$ or $\eta<2$, and ${\cal E}>1$. Finally, we measure the inflow of the matter to the black hole or the accretion rate $(\dot{M}_{\rm acc})$ by integrating the Eq. (\ref{eq-15}) on the event horizon  ($r_{\rm H}=1 + \sqrt{1-a^2}$).

\begin{figure}
\includegraphics[scale=0.30]{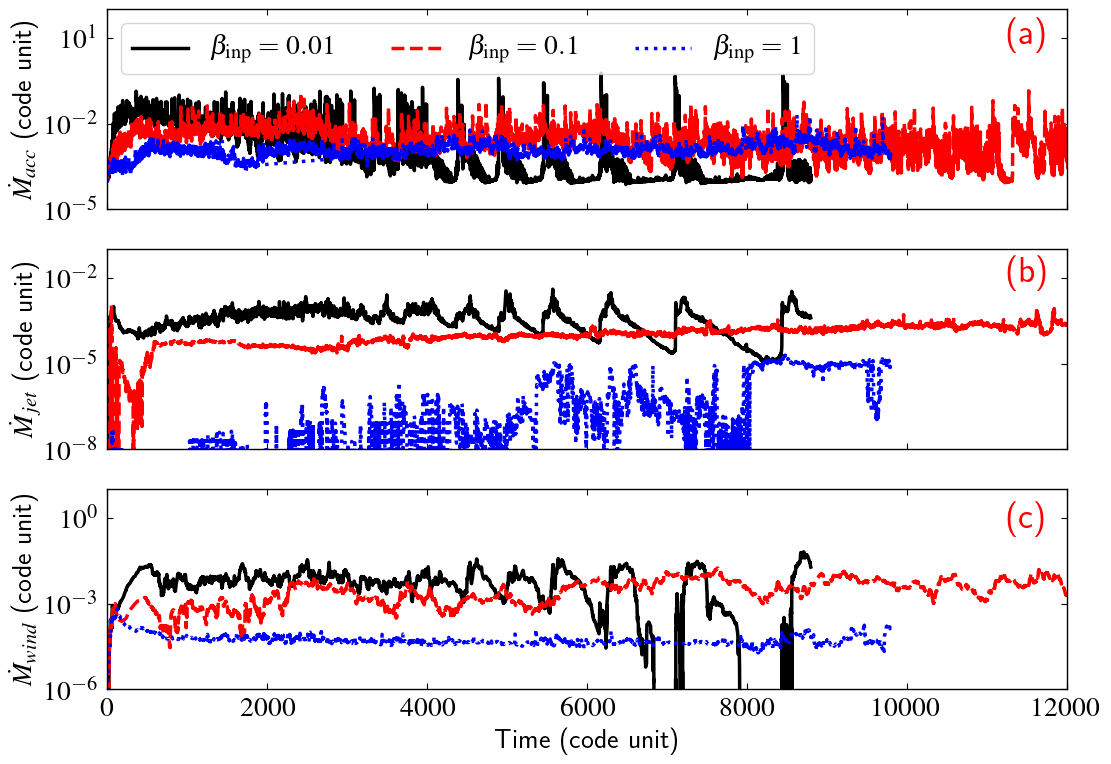}
\caption{Accretion rate $\dot{M}_{\rm acc}$(upper panel),
jet mass flux $\dot{M}_{\rm jet}$ (middle panel), 
and the disc-wind mass flux $\dot{M}_{\rm wind}$ (lower panel) 
as a function of simulation time (in code units). 
Different line styles corresponds to models with different 
initial plasma-$\beta$ 
values as marked on the figure.}
\label{fig-beta-variation}
\end{figure}
\begin{figure}
\includegraphics[scale=0.30]{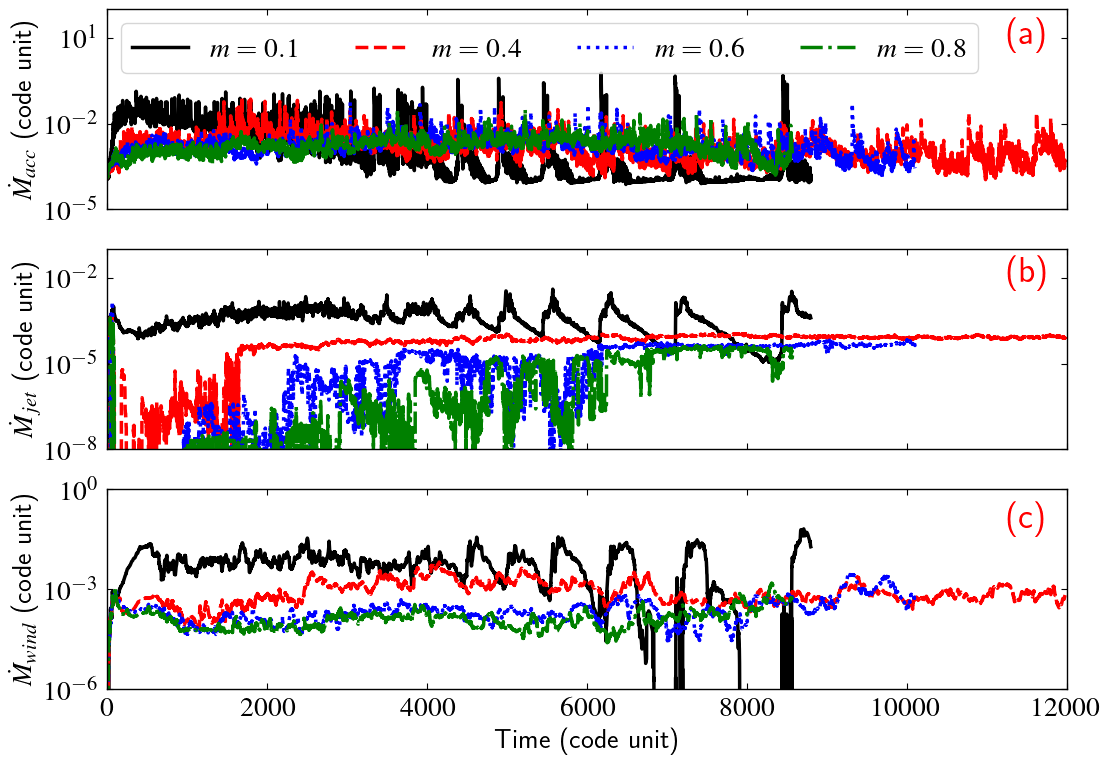}
\caption{Accretion rate $\dot{M}_{\rm acc}$(upper panel), 
jet mass flux $\dot{M}_{\rm jet}$ (middle panel), 
and the disc-wind mass flux $\dot{M}_{\rm wind}$ (lower panel)
as a function of simulation time (in code units). 
Different line styles correspond to models with different 
initial inclination angles $(m)$ as marked on the figure.}
\label{fig-mm-variation}
\end{figure}

A comparison of temporal evolution for
(a) $\dot{M}_{\rm acc}$, 
(b) $\dot{M}_{\rm jet}$, and 
(c) $\dot{M}_{\rm wind}$ 
is shown in fig. (\ref{fig-beta-variation}) for the different models. As there is no inflow of matter initially ($t$ = 0), $\dot{M}_{\rm acc}$ starts evolving from a zero value for all the models. In the accretion disc, the flow is gas pressure dominated plasma-$\beta > 1$. The rotation of the flow generates the toroidal component of the magnetic field $B_{\rm tor}$. The resolved MRI then destabilizes the weakly magnetized rotating matter in the accretion disc and drives turbulence. Further, MRI can amplify the toroidal component of the magnetic field (see \cite{Begelman-Pringle2007}), leading to the turbulent $B_{\rm tor}$ dominated disc wind observed in our simulations (see the turbulent region of Fig. \ref{fig-btorbp}). $B_{\rm tor}$ dominated disc-wind transports the angular momentum along with it, thereby allowing accretion of matter towards the black hole. The temporal evolution of the disc accretion rate (see Fig. \ref{fig-beta-variation}-\ref{fig-mm-variation}) shows an initial increase approaching a quasi-steady state value with time. The time required to reach a quasi-steady state for our models A, B, C, D, E, and F is roughly $t\sim 500, 300, 150, 200, 250,$ and $400$, respectively.
The long-term quasi-steady state behavior implies saturation of the process responsible for accretion.

With the increase of the strength of the magnetic field, MRI is quenched, and the disc tends to get a MAD configuration. In such a situation, the source of turbulence is possibly MRTI (magnetic-Rayleigh-Taylor instabilities) (see section 2.5). In the presence of MRTI, the transport of angular momentum (large scale magnetic stress) is dominantly governed due to strong BP disc-winds. As a result, the quasi-steady accretion rate obtained from the models with a lower value of $\beta_{\rm inp}$ is higher than that of the model with a higher $\beta_{\rm inp}$ parameter. This confirms that the accretion process is magnetically driven, and the rate of accretion is strongly correlated with the strength of the initial magnetic fields. 
\begin{figure}
\includegraphics[scale=0.30]{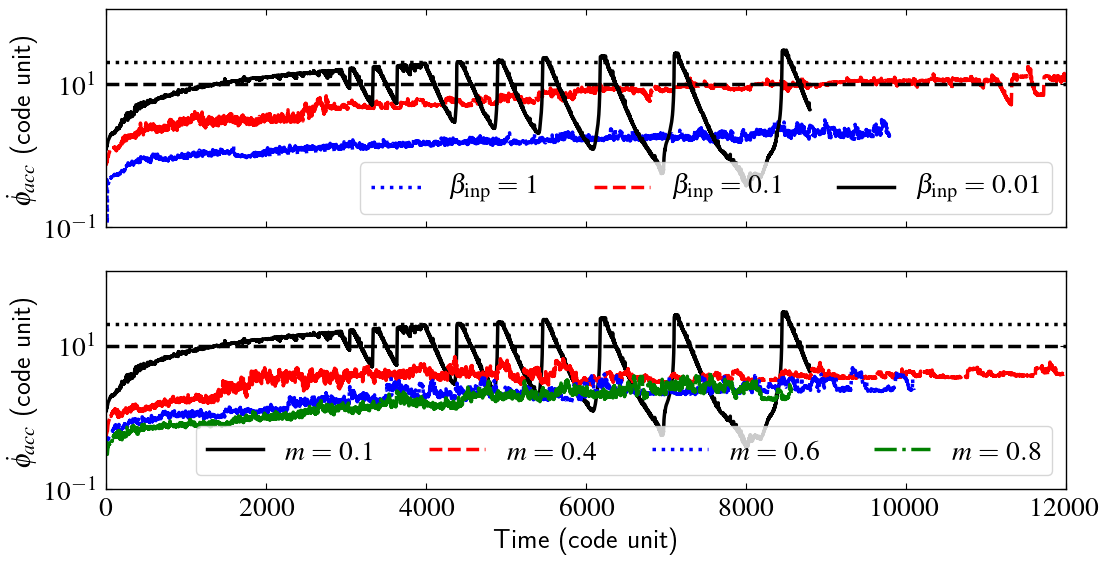}
\caption{The magnetic flux accumulated on the black-hole horizon $(\dot{\phi}_{\rm acc}=\dot{\Phi}_{\rm acc}/\sqrt{<\dot{M}_{\rm acc}>})$
as a function of simulation time (in code units). 
Different line styles correspond to models with different input $\beta_{\rm inp}$ and
initial inclination angles $(m)$ as marked on the figure. Dashed and dotted horizontal lines corresponds to $\dot{\phi}_{\rm acc}=10$ and $20$, respectively.}
\label{fig-mag-flux}
\end{figure}

With the accretion, the magnetic flux accumulates onto the horizon of the black hole. Such accumulation drives accretion disc to MAD configuration, where the magnetic field can reconnect sporadically \citep[and references therein]{Avara-etal2016}.  In Fig. \ref{fig-mag-flux}, we show the temporal evolution of normalized magnetic flux $(\dot{\phi}_{\rm acc}=\dot{\Phi}_{\rm acc}/\sqrt{<\dot{M}_{\rm acc}>})$ for different simulation models. $\dot{\Phi}_{\rm acc}$ is given by \citep{Nathanail-etal2019},
\begin{align}
\dot{\Phi}_{\rm acc}= \frac{1}{2} \int_0^{2\pi} \int_0^\pi \sqrt{-g}|B^r| d\theta d\phi.
\label{eq-16}
\end{align}
Typically normalized magnetic flux value $\dot{\phi}_{\rm acc}\sim 15$ (in the code unit) is categorized as MAD configuration \cite{Tchekhovskoy-etal2011}. From the figure~\ref{fig-mag-flux}, it is clear that model C reaches the MAD configuration at $t\sim3000$ resulting in initiating the oscillatory behaviour of accretion rate. In the oscillating phase, the value of the accretion rate drops down by three orders of magnitude. For simulation runs with higher value of $\beta_{\rm inp}$, we expect disc will take a longer time to accrete enough magnetic flux to transit into a MAD kind of configuration. Indeed, comparing the profiles for model B  $(t\sim 10000)$ with model C  $(t\sim 3000)$, we find the oscillation starts in model B much later than that of model C. However, for other models magnetic flux is much less compared to the MAD limit, and we do not observe any oscillating feature in the accretion rate for model A within the evolution time considered for this study. 

By comparing the $\dot{M}_{\rm wind}$ profiles from Fig. \ref{fig-beta-variation}b, we find that the models with a higher accretion rate show a higher disc-wind rate. The $B_{\rm tor}$ dominated disc-wind can be related to MRI and/or MRTI driven turbulence in the accretion disc. With the increase of magnetic field strength (model A $\rightarrow$ C), the turbulence driven by MRI is suppressed, possibly leading to turbulence driven by MRTI. Additionally, the BP disc-wind also increases with the increase of initial magnetic field strength. Overall, we observe that the quasi-steady value of the disc-wind rate increases with the increase of magnetic field strength (model A $\rightarrow$ C). When the accretion drops down to the minimum in the oscillating phase, the accretion flow vanishes from the launching site of disc-wind. That results in the oscillation of disc-wind rates as well. It is evident from the figure that the disc-winds oscillating features are quite similar to the oscillating features of the accretion rate. 

Fig. \ref{fig-beta-variation}c shows the profiles for the BZ-jet rate for models A, B, and C. We find that the models with higher accretion rate and higher disc-wind rate show a higher BZ-jet rate. With the increase of the magnetic field strength (model A $\rightarrow$ C), the flow starts evolving from a stronger initial poloidal magnetic field. With temporal evolution, the steady state magnetisation $(\sigma)$ increases in the funnel region. Consequently, we observe a faster BZ-jet with the increase of magnetic field strength (see the reddish region of Fig. \ref{fig-lfac}d-f). Therefore, the mass flux rate through the BZ-jet ($\dot{M}_{\rm jet}$) increases with the magnetic field strength (model A $\rightarrow$ C). As the accretion disc moves far from the event horizon in the oscillating phase, the poloidal component of the magnetic field becomes weaker due to reconnection. As a result, the magnetization of the flow in the funnel region also reduces, which leads to drops in the BZ-jet rate. Because of that, the BZ-jet rate also follows similar oscillating features as the accretion rates in Fig. \ref{fig-beta-variation}a.

We compare  $\dot{M}_{\rm acc}$, $\dot{M}_{\rm jet}$, and $\dot{M}_{\rm wind}$ for models C, D, E, and F, in Fig. \ref{fig-mm-variation}. With the decrease of the inclination angle (lowering $m$), the poloidal component of the magnetic field becomes stronger around the equatorial plane. However, the thin-disc remain gas pressure dominated for all the models ($\beta \gg 1$). Consequently, we find that the quasi-steady value of the accretion rate $\dot{M}_{\rm acc}$, BZ-jet rate $\dot{M}_{\rm jet}$, and disc-wind rate $\dot{M}_{\rm wind}$ decreases with the inclination parameter $m$ (Fig. \ref{fig-mm-variation}a-c). The early onset of oscillation in the mass flux rate for stronger magnetic field configuration is also seen in Fig. \ref{fig-mm-variation} with varying of the inclination parameter $m$ (model F $\rightarrow$ C).

\begin{figure}
\includegraphics[scale=0.4]{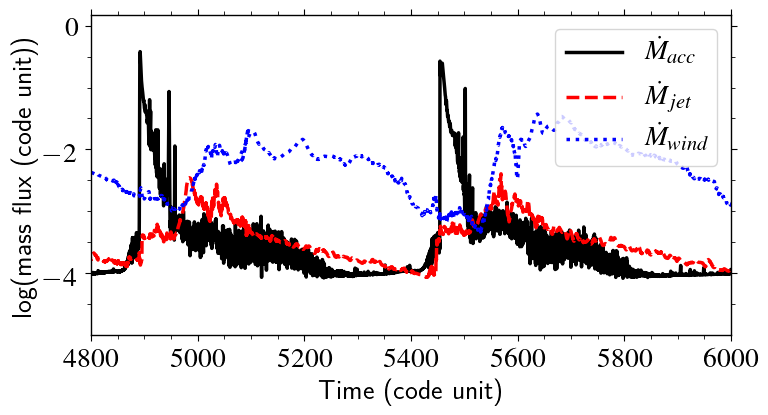}
\caption{Accretion rate $\dot{M}_{\rm acc}$ (black solid), 
jet mass flux $\dot{M}_{\rm jet}$ (red dashed), 
and the disc wind mass flux $\dot{M}_{\rm wind}$ (blue dotted) 
as a function of simulation time (in code units) 
in the oscillating phase of model C.}
\label{fig-comparison}
\end{figure}

In both figures (Fig. \ref{fig-beta-variation}, and Fig. \ref{fig-mm-variation}), we observe that the model with higher accretion rates corresponds to a higher rate of disc-wind and higher BZ-jet. As discussed earlier in this section, the accretion process is driven by the turbulence generated due to the MRI, MRTI, and magneto-centrifugal wind (BP disc-wind) depending on magnetic field properties within the disc. Thus the different micro-physics in the thin accretion disc is very important to study and probably plays a crucial role in deciding the nature and the efficiency of jet and disc-wind.

To understand the relation between the disc-jet-wind in detail, in Fig. \ref{fig-comparison}, we show the temporal evolution of  $\dot{M}_{\rm acc}$, $\dot{M}_{\rm jet}$, and $\dot{M}_{\rm wind}$ together for model C during the oscillating phase. We observe that the accretion rate's peak value is highest, followed by the peak for the disc-wind rate, whereas the peak of the BZ-jet rate is the lowest. We further note that there exists a time lag between the different mass flux rate peaks. BZ-jet and disc-wind launch after accretion flow reach the horizon, and the time taken by the BZ-jet and the disc-wind to reach radius $r=50$ from its launching site is appearing as the time lag in the evolution of mass flux ($\Delta t_{\rm lag}$). Since the BZ-jet has a much faster vertical velocity than that of the disc-wind (the funnel region is darker than that of the disc-wind region of Fig. \ref{fig-fieldlines}-\ref{fig-lfac}), the time lag between the accretion rate peak and BZ-jet rate peak is much smaller than that of the accretion rate peak and disc-wind rate peak. 

From Fig. \ref{fig-comparison}, we can roughly calculate the vertical velocity of the BZ-jet and disc-wind, by estimating $v_{\rm app}=r/\Delta t_{\rm lag}$. In Fig. \ref{fig-comparison}, the values of $\Delta t_{\rm lag}$ for BZ-jet and disc-wind are obtained as $\Delta t_{\rm lag}=100$ and $250$, respectively. Thus, the vertical velocity for BZ-jet and disc-wind at radius $r=50$ is obtain as $v_{\rm app}\sim 0.5c$ and $v_{\rm app}\sim0.2c$, respectively. The velocity $0.5c$ and $0.2c$ corresponds to Lorentz factor, $\Gamma=1.15$ and $\Gamma=1.02$, these values are consistent with values shown in Fig. \ref{fig-lfac}.

\section{Inner disc oscillation}
\begin{figure*}
\includegraphics[scale=0.5]{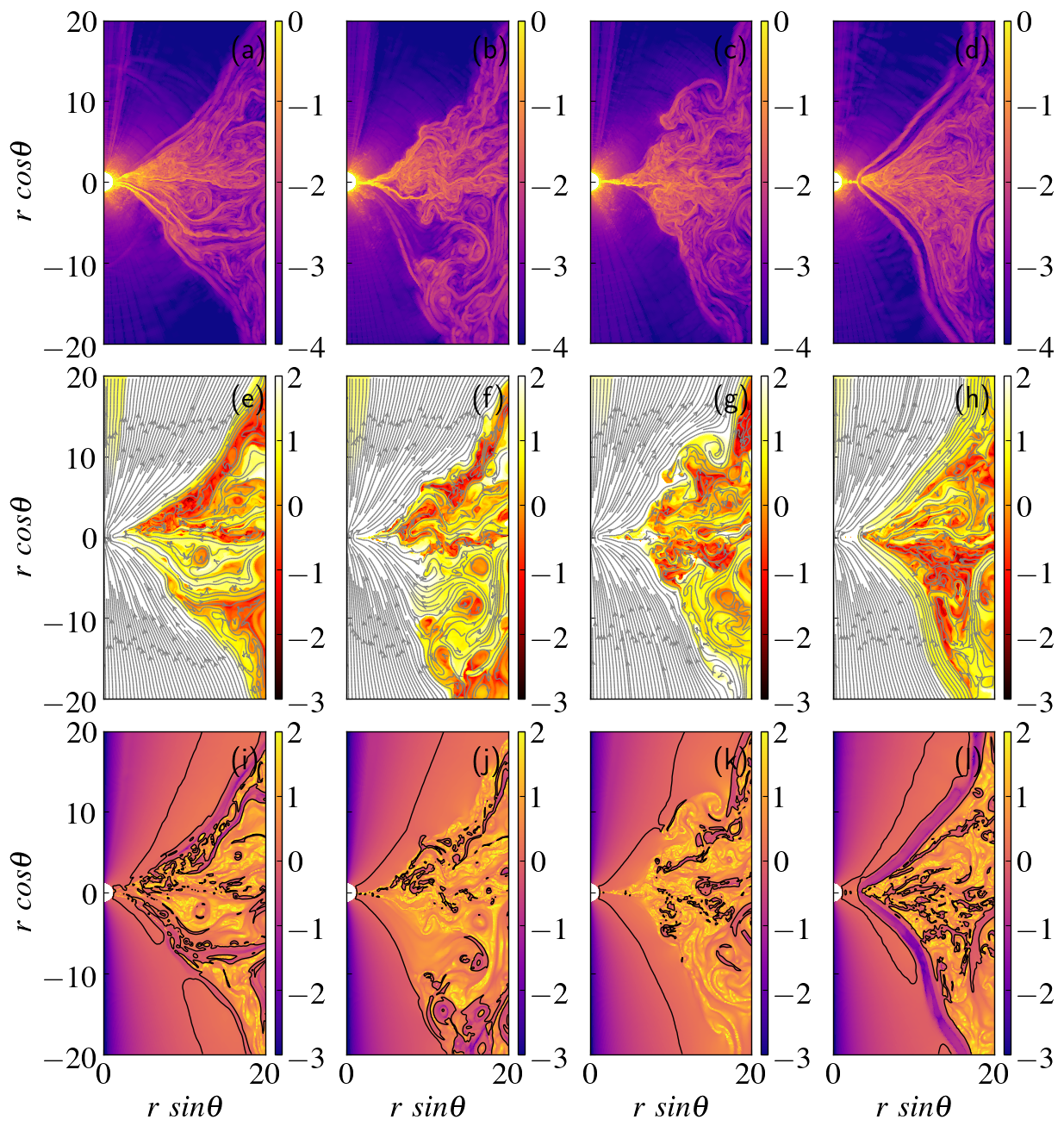}
\caption{For Model C$_{\rm H}:$
logarithmic $|\vec{J}|$ (upper panels), 
logarithmic inverse of the plasma-$\beta$ ($\beta^{-1}=b^2/2p$) with the poloidal field lines in gray color (middle panels), 
and logarithmic $B_{\rm tor}/B_{\rm p}$ (lower panels) in the poloidal 
plane at four different time $t=5575$, $5675$, $5775$ and $5875$ (left to right). The solid lines in the lower panels represent contours of 
$B_{\rm tor}/B_{\rm p}=1$.}
\label{fig-zoom}
\end{figure*}

In our simulation models, we observe that the accretion process starts, allowing the flow to reach up to the event horizon due to the loss of the matter's angular momentum. As time evolves, we notice that the accretion disc truncates and is disconnected from the event horizon (e.g., model C). The truncated disc progressively moves away from the horizon for a brief interval of time when no part of the disc is connected to the event horizon. 

Subsequently, we observe the disc matter fills in the gap generated, and the disc again connects to the event horizon. This process of the inner disc with respect to the event horizon continues, and we observe an oscillating behaviour in the inner part of the accretion disc. The evolution of $\dot{M}_{\rm acc}$, $\dot{M}_{\rm wind}$, and $\dot{M}_{\rm jet}$ also reflects these oscillations (see Fig.~\ref{fig-beta-variation}, and Fig.~\ref{fig-mm-variation}.) 
We also observe formation of plasmoids due to reconnection and turbulent toroidal magnetic field driven winds in this inner region where the disc oscillates. To envisage the role of magnetic fields and current sheets in the disc's oscillating behaviour, we study the temporal evolution of total current density $|\vec{J}| =|\nabla \times \vec{B}|$, the inverse of the plasma-$\beta$ ($\beta^{-1}$), and the ratio between the toroidal and poloidal component of the magnetic field for model C$_{\rm H}$ (Fig. \ref{fig-zoom}).             
In the following, we explain how evolution of these quantities can help us understand the oscillating disc behaviour:  
\begin{enumerate}
    \item As evident from the temporal evolution of the current density ($|\vec{J}|$) (see Fig. \ref{fig-zoom}a-d), we see development of several high current density regions which are morphological thin and typically regarded as current sheets. For example, near the apex point squeezing of opposite polarity magnetic fields result in formation of such sheets. Typically these current sheets are prone to tearing mode instabilities that result in continuous formation of plasmoids . We do observe such plasmoids in our simulation runs particular for model C$_{\rm H}$ (Fig. \ref{fig-plasmoids}).
    
    \item These plasmoids carry the toroidal component of the magnetic fields with them, which increases the magnetic tension force in the funnel region. The developed magnetic tension force plays a vital role in changing the shape of the funnel region, as discussed in section 3 (see the dark blue region of Fig. \ref{fig-density}).
    
    \item The advection of the toroidal magnetic field results in loss of magnetization, allowing the funnel region to expand and thereby increasing the size of the current sheet formed in the under-lying accretion disc.
    
    \item Plasmoids generated due to enhanced reconnection events carry the disc material away, leaving behind a void with newly connected poloidal field lines. 
    
    \item The tension force due to these field lines prevent any further accretion flow of matter, giving rise to a gap between the disc edge and the event horizon (Fig. \ref{fig-zoom}d, h, and j). 
    
    \item At the gap interface, magnetic flux is accumulated with time contributing to the onset of disc-winds and thereby transport the angular momentum. This assists in allowing accretion of matter towards the black hole.  At a time when the accretion of matter over-powers the tension force, it results in a sudden inflow of material filling the gap.
    
    \item Such a sudden inflow further increases the inclination of the poloidal fields, thereby allowing it to reconnect over a larger distance within the disc (Fig. \ref{fig-zoom}e-h).
    
    \item This cycle continues to generate another instance of gaps, which are broader than the previous instance. Eventually, the accretion dominates and forces the matter to inflow towards the black hole resulting in oscillations.  
    
    \item As the gap widens, the contact region of opposite polarity field lines increases. This leads to an increase in the feasibility of reconnection through tearing mode instabilities.
    As time evolves, the reconnection region progressively increases in size with every oscillation resulting in widening the gap separation. Thus, the time period of oscillation also increases.
\end{enumerate}

It is to be noted that the reconnection of field lines and the formation of the plasmoids in our simulation is purely due to the resistivity present in the code due to the finite resolution. \citet{Nathanail-etal2020} have also demonstrated reconnection of field lines and the formation of plasmoids in their simulations using an ideal GRMHD setup. They claim that the qualitative feature of the results will not alter with the inclusion of physical resistivity. We also compare results obtained from model C and C$_{\rm H}$ and find that the qualitative nature of the results remains the same for these two models. Thus, we believe that the formation of plasmoids and the oscillations in the inner part of the accretion disc are qualitatively consistent. However, the quantitative behavior of these phenomena will depend on the physical resistivity present in the flow. Recent simulations by \cite{Vourellis-etal2019, Ripperda-etal2020} have also demonstrated the formation of plasmoids using a resistive GRMHD setup.

\section{Astrophysical context}
\begin{figure}
\centering
\includegraphics[scale=0.38]{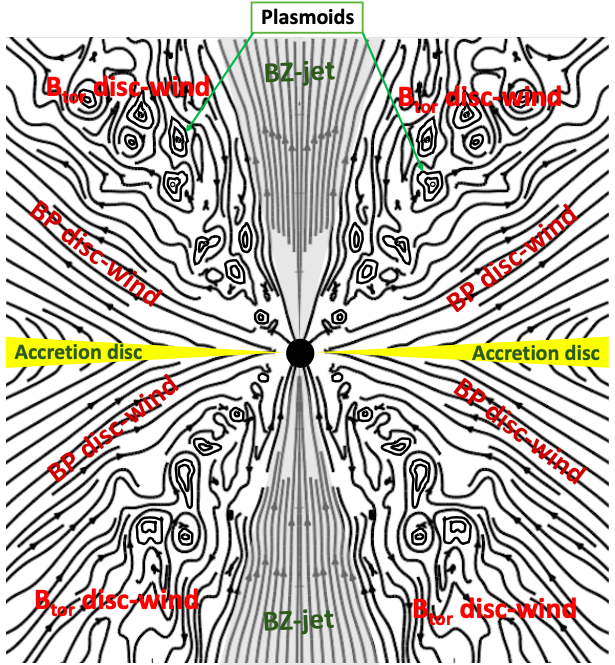}
\caption{Schematic diagram diagram of disc-jet-wind configuration.}
\label{fig-cartoon}
\end{figure}
In Fig. \ref{fig-cartoon}, we present a schematic diagram (toy model) of disc-jet structure based on our simulation results. We observe such structure in our reference model after several tens of inner disc rotation and remain there up to the end of the simulation (several hundreds of inner disc rotation). 

The high-density thin accretion disc occupies the area near the disc-equatorial plane. The region near the rotation axis is occupied by very low density, relativistic, and highly magnetized BZ-jets, where the ratio $B_{\rm tor}/B_{\rm p}\ll 1$. The shape of the BZ-jet is supported by a region with $B_{\rm tor}/B_{\rm p}\gg 1$. We called this region as $B_{\rm tor}$ dominated disc-wind, and the driving mechanism for this disc-wind is the excess toroidal magnetic pressure. $B_{\rm tor}$ dominated disc-wind is turbulent due to the different instabilities present in the thin accretion disc, and we observe plasmoids in this region (see section 5). Further, such a disc-wind could play a very important role in the collimation or the de-collimation of the BZ-jet due to external support.

The $B_{\rm tor}$ dominated disc-wind region is a potential site for turbulent mass loading from the accretion disc \citep{Britzen-etal2017}. The area between the $B_{\rm tor}$ dominated wind and the accretion disc is occupied by BP disc-wind. In this region, the flow around the disc surface has $B_{\rm tor}/B_{\rm p}\lesssim 1$, the driving mechanism of the BP disc-wind is the magneto-centrifugal acceleration as proposed by \cite{Blandford-Payne1982}. The flow leaves the surface of the accretion disc with sub-Alfv\'enic poloidal velocity ($M_{\rm A, p}<1$). As the BP disc-wind moves far from the equatorial plane, the ratio $B_{\rm tor}/B_{\rm p}$ increases, eventually the ratio becomes $B_{\rm tor}/B_{\rm p}\gg 1$ and the flow attains a super-Alfv\'enic poloidal velocity ($M_{\rm A, p}>1$). Thus, both $B_{\rm tor}$ dominated disc-wind and BP disc-wind are super-Alfv\'enic and $B_{\rm tor}/B_{\rm p}\gg 1$ far from the equatorial plane, in such scenario, the flow self-collimates and evolves into a large-scale jet \citep[and references therein]{Fendt2006}.

The toy model proposed above can be applied in several astrophysical contexts. In particular, we want to discuss the implications of our unified disc-jet launching model to outbursts of Black Hole X-ray binaries (BH-XRBs). The BH-XRBs remain in a quiescent state for an extended period, followed by an occasional outburst (e.g., GRO J1655-40, GS1354-64, GX339-4, SWIFTJ1745-26, V-404 Cyg, etc.). During the outburst, BH-XRBs are observed in several prominent spectral states, namely, low-hard state (LHS), hard-intermediate state (HIMS), soft-intermediate state (SIMS), and high-soft state (HSS), respectively \citep{Belloni2010, Belloni-etal2011, Ingram-Motta2019}.

\begin{figure}
\centering
\includegraphics[scale=0.38]{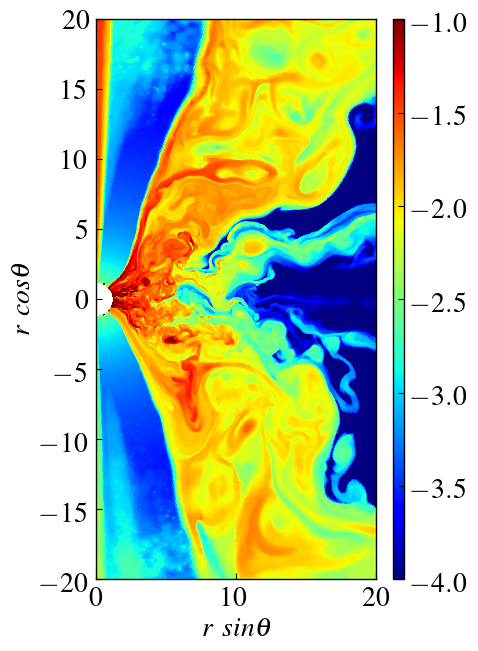}
\caption{Temperature distribution $p/\rho$ in the poloidal plane for model C$_{\rm H}$ at $t=1000$.}
\label{fig-temperature}
\end{figure}

Generally, an outburst starts with LHS, where a steady radio jet is observed \citep{Belloni-Motta2016}.  Hard X-ray dominates the spectrum of the LHS, however, the origin of these hard radiations is still elusive. The plasmoids formed in our simulations can be vital for the abundance of such high-energy non-thermal radiations. In general, electrons are accelerated to relativistic energies due to the direct electric field generated at the reconnection sites. These electrons are possible candidates for the observed non-thermal emissions \citep{Sironi-Spitkovsky2014, Kagan-etal2015, Sironi-etal2015}. A significant portion of dissipated energy from the plasmoids is also responsible for heating the disc-wind resulting in the formation of hot disc-wind corona around the black hole. The distribution of $p/\rho$ in the poloidal plane for model C$_{\rm H}$ at time $t=1000$ is shown in Fig.~\ref{fig-temperature} . The flow in the corona can have temperature as hot as $T\sim 10^{11}$K (One unit of $p/\rho$ corresponds to $T=5.44565\times10^{12}$K of the flow temperature). Such a hot corona around the black hole also contributes to the hard X-ray due to the inverse Comptonization phenomena \citep[etc.]{Haardt-Maraschi1991,Haardt-Maraschi1993, Svensson-Zdziarski1994,Chakrabarti-Titarchuk1995}. 

With time an outburst makes a transition from LHS to hard-intermediate state (HIMS) \citep{Belloni-Motta2016}.
In the HIMS, the radio-activity weakens, and it gives occasional giant radio flares \citep{Fender-etal2009}. Such a situation can be easily correlated to the oscillating phase of our simulation. In the oscillating phase, the inner part of the disc is devoid of matter. Thus, only the outer part of the accretion disc mostly contributes to the emission spectrum, and as a result, the hard radiations from the accretion disc weakens. 

In the oscillating phase, we observe that the BZ-rate ($\dot{M}_{\rm jet}$) shows strong peak values in between two silent periods (Fig. \ref{fig-beta-variation} - Fig. \ref{fig-comparison}). These strong peaks in the ($\dot{M}_{\rm jet}$) can be easily associated with the giant radio flares in the HIMS. We observe that with temporal evolution, our simulation model reaches a steady state, and with further evolution, the simulation reaches an oscillating phase. Thus, we find that our simulation can indicate the mechanism of the transition from LHS to HIMS. Note that our interpretation is only indicative, and the quantitative behavior may vary on the choice of physical resistivity present in the flow (not considered in this work).

\section{Summary and Discussion}
In this work, we have set up a highly resolved, axisymmetric, magnetized thin accretion disc in a GRMHD framework to study the launching of jets, disc-winds and understand their connection with the underlying disc. We have extensively studied the influence of magnetic field strength (plasma- $ \beta $) and the initial inclination angle of the filed lines represented by the parameter $m$. The main results of the present work can be summarized as follows:

\begin{itemize}
\item[(1)] 
From our simulation runs, we can identify three distinct regions: high density accretion disc, slow moving disc-wind, and low density funnel with relativistic velocity. The disc-wind can be further divided into $B_{\rm tor}$ dominated launched due to magnetic pressure gradient, and BP disc-wind driven by magneto-centrifugal acceleration. We also find signatures of active BZ mechanism in launching jet from detailed studies of magnetization, Alfv\'enic Mach number, and the field line structure within the funnel region. These jets are relativistic with Lorentz factors as high as $\gamma\sim10$. 

\item[(2)] The BP disc-wind from the outer parts ($r \gtrsim 10$) and the $B_{\rm tor}$ dominated disc-wind from relatively inner disc regions play an integral role in transporting angular momentum and thereby aiding the accretion process. For all of our runs, we observe the development of turbulence in the accretion disc. Such turbulence can be attributed to magnetic instabilities like MRI, particularly in cases with weak magnetic fields or MRTI in strong field cases.  

\item[(3)] By comparing the results obtained from models with different magnetic field strengths, we find that the strength of the supplied magnetic field plays a major role in the evolution and launching of jets and disc-wind. 

For runs with higher magnetic field strengths - 
    \begin{itemize}
        \item[(i)] we find evidence of a wider jet at the base that is moving with terminal Lorentz factor higher by a factor of $\sim 10$ as compared to runs with the weak field.
        \item[(ii)] we observe that the  quasi-steady value of the accretion rate $(\dot{M}_{\rm acc})$, BZ-rate $(\dot{M}_{\rm jet})$, and disc-wind rate $(\dot{M}_{\rm wind})$ also increases. 
        \item[(iii)] we recognize that the contribution of the BP driven winds increases, while the contribution of $B_{\rm tor}$ dominated wind is reduced and is seen to be launched from a relatively narrow region of the underlying disk. 
        \item[(iv)] we see that jets have a lower degree of collimation as compared to their weak field counterparts primarily due to the loss of external confinement from weaker and less prominent $B_{\rm tor}$ dominated wind. 
    \end{itemize}

\item[(4)] In addition to the strength of the magnetic field, we also observe considerable influence of magnetic field structure in determining the disc, jet, and disc-wind structure.

For runs with higher magnetic field inclination (lower value of $m$) and thereby higher magnetic flux around the disk surface - 
\begin{itemize}
    \item[(i)] we find evidence of a wider jet with a higher Lorentz factor as compared to runs that have more vertical magnetic field lines at the disc-surface. In particular, $\gamma \sim 1.5$ (m=0.8) and $\gamma \sim 10$ (m=0.1).
    \item[(ii)] we observe that BP driven wind is more probable, while for cases with vertical field lines, the main disc-wind launching mechanism is likely to be excess toroidal magnetic pressure. 
\end{itemize}

\item[(5)] For the reference run with m = 0.1 and $\beta_{\rm max} \sim 188$ with a magnetically arrested disc (MAD), we observe the formation of plasmoids due to reconnection close to the black hole. These plasmoids are advected away from the disc along with toroidal magnetic fields. This results in enhancing the magnetic tension from reconnected poloidal field lines in the accretion disc's inner part. 
The accretion process is hindered, and the disc is truncated away from the black hole. 

Such plasmoids formed due to reconnection events in the current sheets could be responsible for the expulsion of magnetic flux and have been associated to understand the flare from Sgr A* \cite[2D GRMHD]{Ripperda-etal2020} \cite[3D GRMHD]{Dexter-etal2020, Porth-etal2021}. Recent 3D ideal GRMHD simulations  \cite{Porth-etal2021} find that the formation of strong current sheets may be hampered due to the presence of non-axisymmetric interchange of instabilities and shearing motion resulting in dissipation of magnetic and kinetic energy. 
These simulations do not form a distinct gap region between the disk and central black hole as seen in our axisymmetric runs but instead show formations of local low density and high magnetization flux bundles that survive for few orbital periods, particularly in MRI suppressed regions.
In general, the development of strong current sheet and MRI suppressed regions will primarily depend on the growth time scales of non-axisymmetric instabilities along with the strength of shearing. 
Further investigations on the interplay of magnetic reconnection and interchange instabilities on dissipating magnetic fields are required using resistive 3D GRMHD simulations.

\item[(6)] Subsequent evolution of flow shows evidence of oscillating behavior where the gap between the disc truncation radius and event horizon increases with time. The impact of oscillating behavior is also evident in all mass outflow and inflow rates, and it becomes more prominent for cases with higher magnetization.  Also, in these cases, the onset of oscillation is much earlier as compared to weak field runs.    
 
\item[(7)] We have put forward a unified disc-jet-wind toy model highlighting the main observations from our simulation runs and have discussed its possible application to BH-XRBs.
The plasmoids and the hot disc-wind corona have the potential to become an efficient tool to understand the flat spectrum observed in LHS (low hard states) of the BH-XRBs. 
From our simulation models, we suggest that the transition from the LHS to HIMS during the outburst of BH-XRBs is possibly governed by the magnetic field present in the accretion flow.

\end{itemize}

In summary, our simulation models show relativistic BZ-jet from the black hole, $B_{\rm tor}$ dominated disc-wind, BP disc-wind from the underlying accretion disc. An overview of the simulation results from different models is presented in table \ref{tab-02}. We find that the models with a stronger and highly inclined magnetic field show prominent BP disc-wind. While models with a weaker and vertical magnetic field, the disc-wind is primarily $B_{\rm tor}$ dominated. 
We qualitatively demonstrate that the magnetic field reconnects at the accretion disc near the black hole horizon leading to the formation of plasmoids. These plasmoids and associated reconnection can serve as ideal sites for particle acceleration and may even contribute to the high energy neutrino emission via hadronic process \citep{Begelman-etal1990,Stecker-etal1991,Alvarez-etal2004,Sironi-etal2015,Inoue-etal2019}. 
These plasmoids also help to advect the toroidal component of the magnetic field and results in the oscillation in the inner part of the accretion disc. 
We discuss the possible astrophysical importance of our models to devise a road-map to understand the dynamics of the spectral state transition in BH-XRBs. 
The simulations presented in this work assume ideal GRMHD, thereby effects associated with reconnection arising from numerical resistivity dependent on grid resolution. In our future work, we will consider relaxing the ideal GRMHD constraint. 

\begin{table*}
\centering
  \begin{tabular}{ c  c  c  c  c  c }
    \hline
    Model & $\beta_{\rm max}$ & MRI status & Possible & Dominant & Collimation degree\\ 
      &   & (see Fig. \ref{fig-qtheta})& instabilities & disc-wind  & of BZ jet\\
    \hline
    \hline
     F & 22657  &  UR & MRI & $B_{\rm tor}$ & high\\
     A & 18844 &  UR &  MRI & $B_{\rm tor}$ & high\\
     E & 13939 &  UR & MRI  &$B_{\rm tor}$  & high\\
     D & 6163 &  R & MRI &  $B_{\rm tor}$  & high\\
     B & 1884 &  R & MRI & $B_{\rm tor}$  & high\\
     C & 188 &  R & MRI + MRTI & $B_{\rm tor}$ + BP  & low\\
     C$_{\rm H}$ & 188 &  HR & MRI + MRTI & $B_{\rm tor}$ + BP & low \\
    \hline
  \end{tabular}
\caption{A summary of results obtained from different simulation models. We show the simulation ID (first column), the initial maximum disk plasma beta (second column), the MRI status of the disk (third column; under-resolved (UR) or highly-resolved (HR)), the potential instabilities present in the disk (4th column), the dominant disk wind driving mechanism (5th column), and the collimation degree of the BZ jet. All features evaluated for the final time step of each simulation.}
\label{tab-02}
\end{table*}
\section*{Acknowledgements}
We would like to thank the anonymous referee for the helpful comments, and constructive remarks on this manuscript.
All simulations were performed on the Max Planck Gesellschaft (MPG) super-computing resources. We would like to thank the financial support from the Max Planck partner group award at Indian Institute Technology of Indore.
\section*{Data availability}
The data underlying this article will be shared on reasonable request to the corresponding author.
\bibliographystyle{mnras}
\bibliography{reference}

\begin{thebibliography}{}
\makeatletter
\relax
\def\mn@urlcharsother{\let\do\@makeother \do\$\do\&\do\#\do\^\do\_\do\%\do\~}
\def\mn@doi{\begingroup\mn@urlcharsother \@ifnextchar [ {\mn@doi@}
  {\mn@doi@[]}}
\def\mn@doi@[#1]#2{\def\@tempa{#1}\ifx\@tempa\@empty \href
  {http://dx.doi.org/#2} {doi:#2}\else \href {http://dx.doi.org/#2} {#1}\fi
  \endgroup}
\def\mn@eprint#1#2{\mn@eprint@#1:#2::\@nil}
\def\mn@eprint@arXiv#1{\href {http://arxiv.org/abs/#1} {{\tt arXiv:#1}}}
\def\mn@eprint@dblp#1{\href {http://dblp.uni-trier.de/rec/bibtex/#1.xml}
  {dblp:#1}}
\def\mn@eprint@#1:#2:#3:#4\@nil{\def\@tempa {#1}\def\@tempb {#2}\def\@tempc
  {#3}\ifx \@tempc \@empty \let \@tempc \@tempb \let \@tempb \@tempa \fi \ifx
  \@tempb \@empty \def\@tempb {arXiv}\fi \@ifundefined
  {mn@eprint@\@tempb}{\@tempb:\@tempc}{\expandafter \expandafter \csname
  mn@eprint@\@tempb\endcsname \expandafter{\@tempc}}}

\bibitem[\protect\citeauthoryear{{Alvarez-Mu{\~n}iz} \&
  {M{\'e}sz{\'a}ros}}{{Alvarez-Mu{\~n}iz} \&
  {M{\'e}sz{\'a}ros}}{2004}]{Alvarez-etal2004}
{Alvarez-Mu{\~n}iz} J.,  {M{\'e}sz{\'a}ros} P.,  2004, \mn@doi [\prd]
  {10.1103/PhysRevD.70.123001}, \href
  {https://ui.adsabs.harvard.edu/abs/2004PhRvD..70l3001A} {70, 123001}

\bibitem[\protect\citeauthoryear{{Avara}, {McKinney}  \& {Reynolds}}{{Avara}
  et~al.}{2016}]{Avara-etal2016}
{Avara} M.~J.,  {McKinney} J.~C.,   {Reynolds} C.~S.,  2016, \mn@doi [\mnras]
  {10.1093/mnras/stw1643}, \href
  {https://ui.adsabs.harvard.edu/abs/2016MNRAS.462..636A} {462, 636}

\bibitem[\protect\citeauthoryear{{Balbus} \& {Hawley}}{{Balbus} \&
  {Hawley}}{1991}]{Balbus-Hawley1991}
{Balbus} S.~A.,  {Hawley} J.~F.,  1991, \mn@doi [\apj] {10.1086/170270}, \href
  {https://ui.adsabs.harvard.edu/abs/1991ApJ...376..214B} {376, 214}

\bibitem[\protect\citeauthoryear{{Balbus} \& {Hawley}}{{Balbus} \&
  {Hawley}}{1998}]{Balbus-Hawley1998}
{Balbus} S.~A.,  {Hawley} J.~F.,  1998, \mn@doi [Reviews of Modern Physics]
  {10.1103/RevModPhys.70.1}, \href
  {https://ui.adsabs.harvard.edu/abs/1998RvMP...70....1B} {70, 1}

\bibitem[\protect\citeauthoryear{{Beckwith}, {Hawley}  \& {Krolik}}{{Beckwith}
  et~al.}{2008}]{Beckwith-etal2008}
{Beckwith} K.,  {Hawley} J.~F.,   {Krolik} J.~H.,  2008, \mn@doi [\apj]
  {10.1086/533492}, \href
  {https://ui.adsabs.harvard.edu/abs/2008ApJ...678.1180B} {678, 1180}

\bibitem[\protect\citeauthoryear{{Beckwith}, {Hawley}  \& {Krolik}}{{Beckwith}
  et~al.}{2009}]{Beckwith-etal2009}
{Beckwith} K.,  {Hawley} J.~F.,   {Krolik} J.~H.,  2009, \mn@doi [\apj]
  {10.1088/0004-637X/707/1/428}, \href
  {https://ui.adsabs.harvard.edu/abs/2009ApJ...707..428B} {707, 428}

\bibitem[\protect\citeauthoryear{{Begelman} \& {Pringle}}{{Begelman} \&
  {Pringle}}{2007}]{Begelman-Pringle2007}
{Begelman} M.~C.,  {Pringle} J.~E.,  2007, \mn@doi [\mnras]
  {10.1111/j.1365-2966.2006.11372.x}, \href
  {https://ui.adsabs.harvard.edu/abs/2007MNRAS.375.1070B} {375, 1070}

\bibitem[\protect\citeauthoryear{{Begelman}, {Rudak}  \& {Sikora}}{{Begelman}
  et~al.}{1990}]{Begelman-etal1990}
{Begelman} M.~C.,  {Rudak} B.,   {Sikora} M.,  1990, \mn@doi [\apj]
  {10.1086/169241}, \href
  {https://ui.adsabs.harvard.edu/abs/1990ApJ...362...38B} {362, 38}

\bibitem[\protect\citeauthoryear{{Belloni}}{{Belloni}}{2010}]{Belloni2010}
{Belloni} T.~M.,  2010, \mn@doi [X-ray Astronomy 2009; Present Status,
  Multi-Wavelength Approach and Future Perspectives] {10.1063/1.3475157}, \href
  {http://adsabs.harvard.edu/abs/2010AIPC.1248..107B} {1248, 107}

\bibitem[\protect\citeauthoryear{{Belloni} \& {Motta}}{{Belloni} \&
  {Motta}}{2016}]{Belloni-Motta2016}
{Belloni} T.~M.,  {Motta} S.~E.,  2016, {Transient Black Hole Binaries}.
p.~61, \mn@doi{10.1007/978-3-662-52859-4_2}

\bibitem[\protect\citeauthoryear{{Belloni}, {Motta}  \&
  {Mu{\~n}oz-Darias}}{{Belloni} et~al.}{2011}]{Belloni-etal2011}
{Belloni} T.~M.,  {Motta} S.~E.,   {Mu{\~n}oz-Darias} T.,  2011, Bulletin of
  the Astronomical Society of India, \href
  {http://adsabs.harvard.edu/abs/2011BASI...39..409B} {39, 409}

\bibitem[\protect\citeauthoryear{{Blandford} \& {Payne}}{{Blandford} \&
  {Payne}}{1982}]{Blandford-Payne1982}
{Blandford} R.~D.,  {Payne} D.~G.,  1982, \mn@doi [\mnras]
  {10.1093/mnras/199.4.883}, \href
  {https://ui.adsabs.harvard.edu/abs/1982MNRAS.199..883B} {199, 883}

\bibitem[\protect\citeauthoryear{{Blandford} \& {Znajek}}{{Blandford} \&
  {Znajek}}{1977}]{Blandford-Znajek1977}
{Blandford} R.~D.,  {Znajek} R.~L.,  1977, \mn@doi [\mnras]
  {10.1093/mnras/179.3.433}, \href
  {https://ui.adsabs.harvard.edu/abs/1977MNRAS.179..433B} {179, 433}

\bibitem[\protect\citeauthoryear{{Boccardi}, {Krichbaum}, {Bach}, {Bremer}  \&
  {Zensus}}{{Boccardi} et~al.}{2016}]{Boccardi-etal2016}
{Boccardi} B.,  {Krichbaum} T.~P.,  {Bach} U.,  {Bremer} M.,   {Zensus} J.~A.,
  2016, \mn@doi [\aap] {10.1051/0004-6361/201628412}, \href
  {https://ui.adsabs.harvard.edu/abs/2016A&A...588L...9B} {588, L9}

\bibitem[\protect\citeauthoryear{{Bonanno} \& {Urpin}}{{Bonanno} \&
  {Urpin}}{2008}]{Bonanno-Urpin2008}
{Bonanno} A.,  {Urpin} V.,  2008, \mn@doi [\aap] {10.1051/0004-6361:20077562},
  \href {https://ui.adsabs.harvard.edu/abs/2008A&A...480...27B} {480, 27}

\bibitem[\protect\citeauthoryear{{Britzen}, {Fendt}, {Eckart}  \&
  {Karas}}{{Britzen} et~al.}{2017}]{Britzen-etal2017}
{Britzen} S.,  {Fendt} C.,  {Eckart} A.,   {Karas} V.,  2017, \mn@doi [\aap]
  {10.1051/0004-6361/201629469}, \href
  {https://ui.adsabs.harvard.edu/abs/2017A&A...601A..52B} {601, A52}

\bibitem[\protect\citeauthoryear{{Bronzwaer}, {Davelaar}, {Younsi},
  {Mo{\'s}cibrodzka}, {Falcke}, {Kramer}  \& {Rezzolla}}{{Bronzwaer}
  et~al.}{2018}]{Bronzwaer-etal2018}
{Bronzwaer} T.,  {Davelaar} J.,  {Younsi} Z.,  {Mo{\'s}cibrodzka} M.,  {Falcke}
  H.,  {Kramer} M.,   {Rezzolla} L.,  2018, \mn@doi [\aap]
  {10.1051/0004-6361/201732149}, \href
  {https://ui.adsabs.harvard.edu/abs/2018A&A...613A...2B} {613, A2}

\bibitem[\protect\citeauthoryear{{Bronzwaer}, {Younsi}, {Davelaar}  \&
  {Falcke}}{{Bronzwaer} et~al.}{2020}]{Bronzwaer-etal2020}
{Bronzwaer} T.,  {Younsi} Z.,  {Davelaar} J.,   {Falcke} H.,  2020, \mn@doi
  [\aap] {10.1051/0004-6361/202038573}, \href
  {https://ui.adsabs.harvard.edu/abs/2020A&A...641A.126B} {641, A126}

\bibitem[\protect\citeauthoryear{{Chakrabarti} \& {Titarchuk}}{{Chakrabarti} \&
  {Titarchuk}}{1995}]{Chakrabarti-Titarchuk1995}
{Chakrabarti} S.,  {Titarchuk} L.~G.,  1995, \mn@doi [\apj] {10.1086/176610},
  \href {https://ui.adsabs.harvard.edu/abs/1995ApJ...455..623C} {455, 623}

\bibitem[\protect\citeauthoryear{{Chatterjee} et~al.,}{{Chatterjee}
  et~al.}{2020}]{Chatterjee-etal2020}
{Chatterjee} K.,  et~al., 2020, \mn@doi [\mnras] {10.1093/mnras/staa2718},
  \href {https://ui.adsabs.harvard.edu/abs/2020MNRAS.499..362C} {499, 362}

\bibitem[\protect\citeauthoryear{{Davis} \& {Tchekhovskoy}}{{Davis} \&
  {Tchekhovskoy}}{2020}]{Davis-Tchekhovskoy2020}
{Davis} S.~W.,  {Tchekhovskoy} A.,  2020, \mn@doi [\araa]
  {10.1146/annurev-astro-081817-051905}, \href
  {https://ui.adsabs.harvard.edu/abs/2020ARA&A..5881817D} {58, annurev}

\bibitem[\protect\citeauthoryear{{De Villiers} \& {Hawley}}{{De Villiers} \&
  {Hawley}}{2003}]{DeVilliers-etal2003}
{De Villiers} J.-P.,  {Hawley} J.~F.,  2003, \mn@doi [\apj] {10.1086/375866},
  \href {https://ui.adsabs.harvard.edu/abs/2003ApJ...592.1060D} {592, 1060}

\bibitem[\protect\citeauthoryear{{Del Zanna}, {Zanotti}, {Bucciantini}  \&
  {Londrillo}}{{Del Zanna} et~al.}{2007}]{DelZanna-etal2007}
{Del Zanna} L.,  {Zanotti} O.,  {Bucciantini} N.,   {Londrillo} P.,  2007,
  \mn@doi [\aap] {10.1051/0004-6361:20077093}, \href
  {https://ui.adsabs.harvard.edu/abs/2007A&A...473...11D} {473, 11}

\bibitem[\protect\citeauthoryear{{Del Zanna}, {Papini}, {Landi}, {Bugli}  \&
  {Bucciantini}}{{Del Zanna} et~al.}{2016}]{DelZanna-etal2016}
{Del Zanna} L.,  {Papini} E.,  {Landi} S.,  {Bugli} M.,   {Bucciantini} N.,
  2016, \mn@doi [\mnras] {10.1093/mnras/stw1242}, \href
  {https://ui.adsabs.harvard.edu/abs/2016MNRAS.460.3753D} {460, 3753}

\bibitem[\protect\citeauthoryear{{Dexter}}{{Dexter}}{2016}]{Dexter2016}
{Dexter} J.,  2016, \mn@doi [\mnras] {10.1093/mnras/stw1526}, \href
  {https://ui.adsabs.harvard.edu/abs/2016MNRAS.462..115D} {462, 115}

\bibitem[\protect\citeauthoryear{{Dexter} et~al.,}{{Dexter}
  et~al.}{2020}]{Dexter-etal2020}
{Dexter} J.,  et~al., 2020, \mn@doi [\mnras] {10.1093/mnras/staa2288}, \href
  {https://ui.adsabs.harvard.edu/abs/2020MNRAS.497.4999D} {497, 4999}

\bibitem[\protect\citeauthoryear{{Dihingia}, {Das}  \& {Mandal}}{{Dihingia}
  et~al.}{2018}]{Dihingia-etal2018}
{Dihingia} I.~K.,  {Das} S.,   {Mandal} S.,  2018, \mn@doi [\mnras]
  {10.1093/mnras/stx3269}, \href
  {https://ui.adsabs.harvard.edu/abs/2018MNRAS.475.2164D} {475, 2164}

\bibitem[\protect\citeauthoryear{{Dihingia}, {Das}, {Prabhakar}  \& {Mand
  al}}{{Dihingia} et~al.}{2020}]{Dihingia-etal2020}
{Dihingia} I.~K.,  {Das} S.,  {Prabhakar} G.,   {Mand al} S.,  2020, \mn@doi
  [\mnras] {10.1093/mnras/staa1687}, \href
  {https://ui.adsabs.harvard.edu/abs/2020MNRAS.496.3043D} {496, 3043}

\bibitem[\protect\citeauthoryear{{Doeleman} et~al.,}{{Doeleman}
  et~al.}{2012}]{Doeleman-etal2012}
{Doeleman} S.~S.,  et~al., 2012, \mn@doi [Science] {10.1126/science.1224768},
  \href {https://ui.adsabs.harvard.edu/abs/2012Sci...338..355D} {338, 355}

\bibitem[\protect\citeauthoryear{{Fender} \& {Gallo}}{{Fender} \&
  {Gallo}}{2014}]{Fender-Gallo2014}
{Fender} R.,  {Gallo} E.,  2014, \mn@doi [\ssr] {10.1007/s11214-014-0069-z},
  \href {https://ui.adsabs.harvard.edu/abs/2014SSRv..183..323F} {183, 323}

\bibitem[\protect\citeauthoryear{{Fender}, {Homan}  \& {Belloni}}{{Fender}
  et~al.}{2009}]{Fender-etal2009}
{Fender} R.~P.,  {Homan} J.,   {Belloni} T.~M.,  2009, \mn@doi [\mnras]
  {10.1111/j.1365-2966.2009.14841.x}, \href
  {https://ui.adsabs.harvard.edu/abs/2009MNRAS.396.1370F} {396, 1370}

\bibitem[\protect\citeauthoryear{{Fendt}}{{Fendt}}{2006}]{Fendt2006}
{Fendt} C.,  2006, \mn@doi [\apj] {10.1086/507976}, \href
  {https://ui.adsabs.harvard.edu/abs/2006ApJ...651..272F} {651, 272}

\bibitem[\protect\citeauthoryear{Fernandes, Garcia  \& Lima}{Fernandes
  et~al.}{2012}]{Fernandes-etal2012}
Fernandes A.~J.,  Garcia P.~J.,   Lima J.~J.,  2012, Jets in Young Stellar
  Objects: Theory and Observations.
Springer Science \& Business Media

\bibitem[\protect\citeauthoryear{{Haardt} \& {Maraschi}}{{Haardt} \&
  {Maraschi}}{1991}]{Haardt-Maraschi1991}
{Haardt} F.,  {Maraschi} L.,  1991, \mn@doi [\apjl] {10.1086/186171}, \href
  {https://ui.adsabs.harvard.edu/abs/1991ApJ...380L..51H} {380, L51}

\bibitem[\protect\citeauthoryear{{Haardt} \& {Maraschi}}{{Haardt} \&
  {Maraschi}}{1993}]{Haardt-Maraschi1993}
{Haardt} F.,  {Maraschi} L.,  1993, \mn@doi [\apj] {10.1086/173020}, \href
  {https://ui.adsabs.harvard.edu/abs/1993ApJ...413..507H} {413, 507}

\bibitem[\protect\citeauthoryear{{Han}}{{Han}}{2017}]{Han2017}
{Han} J.~L.,  2017, \mn@doi [\araa] {10.1146/annurev-astro-091916-055221},
  \href {https://ui.adsabs.harvard.edu/abs/2017ARA&A..55..111H} {55, 111}

\bibitem[\protect\citeauthoryear{{Hardee}, {Mizuno}  \& {Nishikawa}}{{Hardee}
  et~al.}{2007}]{Hardee-etal2007}
{Hardee} P.,  {Mizuno} Y.,   {Nishikawa} K.-I.,  2007, \mn@doi [\apss]
  {10.1007/s10509-007-9529-1}, \href
  {https://ui.adsabs.harvard.edu/abs/2007Ap&SS.311..281H} {311, 281}

\bibitem[\protect\citeauthoryear{{Hawley} \& {Krolik}}{{Hawley} \&
  {Krolik}}{2002}]{Hawley-etal2002}
{Hawley} J.~F.,  {Krolik} J.~H.,  2002, \mn@doi [\apj] {10.1086/338059}, \href
  {https://ui.adsabs.harvard.edu/abs/2002ApJ...566..164H} {566, 164}

\bibitem[\protect\citeauthoryear{{Igumenshchev}}{{Igumenshchev}}{2008}]{Igumenshchev2008}
{Igumenshchev} I.~V.,  2008, \mn@doi [\apj] {10.1086/529025}, \href
  {https://ui.adsabs.harvard.edu/abs/2008ApJ...677..317I} {677, 317}

\bibitem[\protect\citeauthoryear{{Inda-Koide}, {Koide}  \&
  {Morino}}{{Inda-Koide} et~al.}{2019}]{Koide-etal2019}
{Inda-Koide} M.,  {Koide} S.,   {Morino} R.,  2019, \mn@doi [\apj]
  {10.3847/1538-4357/ab345f}, \href
  {https://ui.adsabs.harvard.edu/abs/2019ApJ...883...69I} {883, 69}

\bibitem[\protect\citeauthoryear{{Ingram} \& {Motta}}{{Ingram} \&
  {Motta}}{2019}]{Ingram-Motta2019}
{Ingram} A.~R.,  {Motta} S.~E.,  2019, \mn@doi [\nar]
  {10.1016/j.newar.2020.101524}, \href
  {https://ui.adsabs.harvard.edu/abs/2019NewAR..8501524I} {85, 101524}

\bibitem[\protect\citeauthoryear{{Inoue}, {Khangulyan}, {Inoue}  \&
  {Doi}}{{Inoue} et~al.}{2019}]{Inoue-etal2019}
{Inoue} Y.,  {Khangulyan} D.,  {Inoue} S.,   {Doi} A.,  2019, \mn@doi [\apj]
  {10.3847/1538-4357/ab2715}, \href
  {https://ui.adsabs.harvard.edu/abs/2019ApJ...880...40I} {880, 40}

\bibitem[\protect\citeauthoryear{{Kadowaki}, {De Gouveia Dal Pino}  \&
  {Stone}}{{Kadowaki} et~al.}{2018}]{Kadowaki-etal2018}
{Kadowaki} L. H.~S.,  {De Gouveia Dal Pino} E.~M.,   {Stone} J.~M.,  2018,
  \mn@doi [\apj] {10.3847/1538-4357/aad4ff}, \href
  {https://ui.adsabs.harvard.edu/abs/2018ApJ...864...52K} {864, 52}

\bibitem[\protect\citeauthoryear{{Kagan}, {Sironi}, {Cerutti}  \&
  {Giannios}}{{Kagan} et~al.}{2015}]{Kagan-etal2015}
{Kagan} D.,  {Sironi} L.,  {Cerutti} B.,   {Giannios} D.,  2015, \mn@doi [\ssr]
  {10.1007/s11214-014-0132-9}, \href
  {https://ui.adsabs.harvard.edu/abs/2015SSRv..191..545K} {191, 545}

\bibitem[\protect\citeauthoryear{{Koide}, {Shibata}  \& {Kudoh}}{{Koide}
  et~al.}{1999}]{Koide-etal1999}
{Koide} S.,  {Shibata} K.,   {Kudoh} T.,  1999, \mn@doi [\apj]
  {10.1086/307667}, \href
  {https://ui.adsabs.harvard.edu/abs/1999ApJ...522..727K} {522, 727}

\bibitem[\protect\citeauthoryear{{Komissarov}}{{Komissarov}}{2004}]{Komissarov2004}
{Komissarov} S.~S.,  2004, \mn@doi [\mnras] {10.1111/j.1365-2966.2004.07598.x},
  \href {https://ui.adsabs.harvard.edu/abs/2004MNRAS.350..427K} {350, 427}

\bibitem[\protect\citeauthoryear{{Komissarov} \& {Barkov}}{{Komissarov} \&
  {Barkov}}{2009}]{Komissarov-Barkov2009}
{Komissarov} S.~S.,  {Barkov} M.~V.,  2009, \mn@doi [\mnras]
  {10.1111/j.1365-2966.2009.14831.x}, \href
  {https://ui.adsabs.harvard.edu/abs/2009MNRAS.397.1153K} {397, 1153}

\bibitem[\protect\citeauthoryear{{Marshall}, {Avara}  \& {McKinney}}{{Marshall}
  et~al.}{2018}]{Marshall-etal2018}
{Marshall} M.~D.,  {Avara} M.~J.,   {McKinney} J.~C.,  2018, \mn@doi [\mnras]
  {10.1093/mnras/sty1184}, \href
  {https://ui.adsabs.harvard.edu/abs/2018MNRAS.478.1837M} {478, 1837}

\bibitem[\protect\citeauthoryear{{Matsumoto}, {Kato}, {Fukue}  \&
  {Okazaki}}{{Matsumoto} et~al.}{1984}]{Matsumoto-etal1984}
{Matsumoto} R.,  {Kato} S.,  {Fukue} J.,   {Okazaki} A.~T.,  1984, \pasj, \href
  {https://ui.adsabs.harvard.edu/abs/1984PASJ...36...71M} {36, 71}

\bibitem[\protect\citeauthoryear{{McClintock}, {Shafee}, {Narayan},
  {Remillard}, {Davis}  \& {Li}}{{McClintock}
  et~al.}{2006}]{McClintock-etal2006}
{McClintock} J.~E.,  {Shafee} R.,  {Narayan} R.,  {Remillard} R.~A.,  {Davis}
  S.~W.,   {Li} L.-X.,  2006, \mn@doi [\apj] {10.1086/508457}, \href
  {https://ui.adsabs.harvard.edu/abs/2006ApJ...652..518M} {652, 518}

\bibitem[\protect\citeauthoryear{{McKinney} \& {Gammie}}{{McKinney} \&
  {Gammie}}{2004}]{McKinney-Gammie2004}
{McKinney} J.~C.,  {Gammie} C.~F.,  2004, \mn@doi [\apj] {10.1086/422244},
  \href {https://ui.adsabs.harvard.edu/abs/2004ApJ...611..977M} {611, 977}

\bibitem[\protect\citeauthoryear{{McKinney}, {Tchekhovskoy}  \&
  {Blandford}}{{McKinney} et~al.}{2012}]{McKinney-etal2012}
{McKinney} J.~C.,  {Tchekhovskoy} A.,   {Blandford} R.~D.,  2012, \mn@doi
  [\mnras] {10.1111/j.1365-2966.2012.21074.x}, \href
  {https://ui.adsabs.harvard.edu/abs/2012MNRAS.423.3083M} {423, 3083}

\bibitem[\protect\citeauthoryear{{McKinney}, {Tchekhovskoy}, {Sadowski}  \&
  {Narayan}}{{McKinney} et~al.}{2014}]{McKinney-etal2014}
{McKinney} J.~C.,  {Tchekhovskoy} A.,  {Sadowski} A.,   {Narayan} R.,  2014,
  \mn@doi [\mnras] {10.1093/mnras/stu762}, \href
  {https://ui.adsabs.harvard.edu/abs/2014MNRAS.441.3177M} {441, 3177}

\bibitem[\protect\citeauthoryear{{Misner}, {Thorne}  \& {Wheeler}}{{Misner}
  et~al.}{1973}]{Misner-etal1973}
{Misner} C.~W.,  {Thorne} K.~S.,   {Wheeler} J.~A.,  1973, {Gravitation}

\bibitem[\protect\citeauthoryear{{Mo{\'s}cibrodzka}}{{Mo{\'s}cibrodzka}}{2020}]{Moscibrodzka2020}
{Mo{\'s}cibrodzka} M.,  2020, \mn@doi [\mnras] {10.1093/mnras/stz3329}, \href
  {https://ui.adsabs.harvard.edu/abs/2020MNRAS.491.4807M} {491, 4807}

\bibitem[\protect\citeauthoryear{{Nakamura} et~al.,}{{Nakamura}
  et~al.}{2018}]{Nakamura-etal2018}
{Nakamura} M.,  et~al., 2018, \mn@doi [\apj] {10.3847/1538-4357/aaeb2d}, \href
  {https://ui.adsabs.harvard.edu/abs/2018ApJ...868..146N} {868, 146}

\bibitem[\protect\citeauthoryear{{Narayan} \& {McClintock}}{{Narayan} \&
  {McClintock}}{2012}]{Narayan-McClintock2012}
{Narayan} R.,  {McClintock} J.~E.,  2012, \mn@doi [\mnras]
  {10.1111/j.1745-3933.2011.01181.x}, \href
  {https://ui.adsabs.harvard.edu/abs/2012MNRAS.419L..69N} {419, L69}

\bibitem[\protect\citeauthoryear{{Narayan}, {Igumenshchev}  \&
  {Abramowicz}}{{Narayan} et~al.}{2003}]{Narayan-etal2003}
{Narayan} R.,  {Igumenshchev} I.~V.,   {Abramowicz} M.~A.,  2003, \mn@doi
  [\pasj] {10.1093/pasj/55.6.L69}, \href
  {https://ui.adsabs.harvard.edu/abs/2003PASJ...55L..69N} {55, L69}

\bibitem[\protect\citeauthoryear{{Nathanail}, {Porth}  \&
  {Rezzolla}}{{Nathanail} et~al.}{2019}]{Nathanail-etal2019}
{Nathanail} A.,  {Porth} O.,   {Rezzolla} L.,  2019, \mn@doi [\apjl]
  {10.3847/2041-8213/aaf73a}, \href
  {https://ui.adsabs.harvard.edu/abs/2019ApJ...870L..20N} {870, L20}

\bibitem[\protect\citeauthoryear{{Nathanail}, {Fromm}, {Porth}, {Olivares},
  {Younsi}, {Mizuno}  \& {Rezzolla}}{{Nathanail}
  et~al.}{2020}]{Nathanail-etal2020}
{Nathanail} A.,  {Fromm} C.~M.,  {Porth} O.,  {Olivares} H.,  {Younsi} Z.,
  {Mizuno} Y.,   {Rezzolla} L.,  2020, \mn@doi [\mnras]
  {10.1093/mnras/staa1165}, \href
  {https://ui.adsabs.harvard.edu/abs/2020MNRAS.495.1549N} {495, 1549}

\bibitem[\protect\citeauthoryear{{Noble}, {Krolik}  \& {Hawley}}{{Noble}
  et~al.}{2009}]{Noble-etal2009}
{Noble} S.~C.,  {Krolik} J.~H.,   {Hawley} J.~F.,  2009, \mn@doi [\apj]
  {10.1088/0004-637X/692/1/411}, \href
  {https://ui.adsabs.harvard.edu/abs/2009ApJ...692..411N} {692, 411}

\bibitem[\protect\citeauthoryear{{Noble}, {Krolik}, {Schnittman}  \&
  {Hawley}}{{Noble} et~al.}{2011}]{Noble-etal2011}
{Noble} S.~C.,  {Krolik} J.~H.,  {Schnittman} J.~D.,   {Hawley} J.~F.,  2011,
  \mn@doi [\apj] {10.1088/0004-637X/743/2/115}, \href
  {https://ui.adsabs.harvard.edu/abs/2011ApJ...743..115N} {743, 115}

\bibitem[\protect\citeauthoryear{{Novikov} \& {Thorne}}{{Novikov} \&
  {Thorne}}{1973}]{Novikov-Thorne1973}
{Novikov} I.~D.,  {Thorne} K.~S.,  1973, in Black Holes (Les Astres Occlus). pp
  343--450

\bibitem[\protect\citeauthoryear{Oda, Machida, Nakamura  \& Matsumoto}{Oda
  et~al.}{2007}]{Oda-etal2007}
Oda H.,  Machida M.,  Nakamura K.~E.,   Matsumoto R.,  2007, Publications of
  the Astronomical Society of Japan, 59, 457

\bibitem[\protect\citeauthoryear{Oda, Machida, Nakamura  \& Matsumoto}{Oda
  et~al.}{2010}]{Oda-etal2010}
Oda H.,  Machida M.,  Nakamura K.~E.,   Matsumoto R.,  2010, The Astrophysical
  Journal, 712, 639

\bibitem[\protect\citeauthoryear{Oda, Machida, Nakamura, Matsumoto  \&
  Narayan}{Oda et~al.}{2012}]{Oda-etal2012}
Oda H.,  Machida M.,  Nakamura K.~E.,  Matsumoto R.,   Narayan R.,  2012,
  Publications of the Astronomical Society of Japan, 64, 15

\bibitem[\protect\citeauthoryear{{Olivares}, {Porth}, {Davelaar}, {Most},
  {Fromm}, {Mizuno}, {Younsi}  \& {Rezzolla}}{{Olivares}
  et~al.}{2019}]{Olivares-etal2019}
{Olivares} H.,  {Porth} O.,  {Davelaar} J.,  {Most} E.~R.,  {Fromm} C.~M.,
  {Mizuno} Y.,  {Younsi} Z.,   {Rezzolla} L.,  2019, \mn@doi [\aap]
  {10.1051/0004-6361/201935559}, \href
  {https://ui.adsabs.harvard.edu/abs/2019A&A...629A..61O} {629, A61}

\bibitem[\protect\citeauthoryear{{Paczynski} \& {Bisnovatyi-Kogan}}{{Paczynski}
  \& {Bisnovatyi-Kogan}}{1981}]{Paczynski-Kogan1981}
{Paczynski} B.,  {Bisnovatyi-Kogan} G.,  1981, \actaa, \href
  {https://ui.adsabs.harvard.edu/abs/1981AcA....31..283P} {31, 283}

\bibitem[\protect\citeauthoryear{{Page} \& {Thorne}}{{Page} \&
  {Thorne}}{1974}]{Page-Thorne1974}
{Page} D.~N.,  {Thorne} K.~S.,  1974, \mn@doi [\apj] {10.1086/152990}, \href
  {https://ui.adsabs.harvard.edu/abs/1974ApJ...191..499P} {191, 499}

\bibitem[\protect\citeauthoryear{{Peitz} \& {Appl}}{{Peitz} \&
  {Appl}}{1997}]{Peitz-Appl1997}
{Peitz} J.,  {Appl} S.,  1997, \mn@doi [Mon. Not. Roy. Astron. Soc.]
  {10.1093/mnras/286.3.681}, \href
  {http://adsabs.harvard.edu/abs/1997MNRAS.286..681P} {286, 681}

\bibitem[\protect\citeauthoryear{{Penna}, {McKinney}, {Narayan},
  {Tchekhovskoy}, {Shafee}  \& {McClintock}}{{Penna}
  et~al.}{2010}]{Penna-etal2010}
{Penna} R.~F.,  {McKinney} J.~C.,  {Narayan} R.,  {Tchekhovskoy} A.,  {Shafee}
  R.,   {McClintock} J.~E.,  2010, \mn@doi [\mnras]
  {10.1111/j.1365-2966.2010.17170.x}, \href
  {https://ui.adsabs.harvard.edu/abs/2010MNRAS.408..752P} {408, 752}

\bibitem[\protect\citeauthoryear{{Porth}, {Olivares}, {Mizuno}, {Younsi},
  {Rezzolla}, {Moscibrodzka}, {Falcke}  \& {Kramer}}{{Porth}
  et~al.}{2017}]{Porth-etal2017}
{Porth} O.,  {Olivares} H.,  {Mizuno} Y.,  {Younsi} Z.,  {Rezzolla} L.,
  {Moscibrodzka} M.,  {Falcke} H.,   {Kramer} M.,  2017, \mn@doi [Computational
  Astrophysics and Cosmology] {10.1186/s40668-017-0020-2}, \href
  {https://ui.adsabs.harvard.edu/abs/2017ComAC...4....1P} {4, 1}

\bibitem[\protect\citeauthoryear{{Porth} et~al.,}{{Porth}
  et~al.}{2019}]{Porth-etal2019}
{Porth} O.,  et~al., 2019, \mn@doi [\apjs] {10.3847/1538-4365/ab29fd}, \href
  {https://ui.adsabs.harvard.edu/abs/2019ApJS..243...26P} {243, 26}

\bibitem[\protect\citeauthoryear{{Porth}, {Mizuno}, {Younsi}  \&
  {Fromm}}{{Porth} et~al.}{2021}]{Porth-etal2021}
{Porth} O.,  {Mizuno} Y.,  {Younsi} Z.,   {Fromm} C.~M.,  2021, \mn@doi
  [\mnras] {10.1093/mnras/stab163}, \href
  {https://ui.adsabs.harvard.edu/abs/2021MNRAS.502.2023P} {502, 2023}

\bibitem[\protect\citeauthoryear{{Pucci} \& {Velli}}{{Pucci} \&
  {Velli}}{2014}]{Pucci-Velli2014}
{Pucci} F.,  {Velli} M.,  2014, \mn@doi [\apjl] {10.1088/2041-8205/780/2/L19},
  \href {https://ui.adsabs.harvard.edu/abs/2014ApJ...780L..19P} {780, L19}

\bibitem[\protect\citeauthoryear{{Qian}, {Fendt}, {Noble}  \& {Bugli}}{{Qian}
  et~al.}{2017}]{Qian-etal2017}
{Qian} Q.,  {Fendt} C.,  {Noble} S.,   {Bugli} M.,  2017, \mn@doi [\apj]
  {10.3847/1538-4357/834/1/29}, \href
  {https://ui.adsabs.harvard.edu/abs/2017ApJ...834...29Q} {834, 29}

\bibitem[\protect\citeauthoryear{{Qian}, {Fendt}  \& {Vourellis}}{{Qian}
  et~al.}{2018}]{Qian-etal2018}
{Qian} Q.,  {Fendt} C.,   {Vourellis} C.,  2018, \mn@doi [\apj]
  {10.3847/1538-4357/aabd36}, \href
  {https://ui.adsabs.harvard.edu/abs/2018ApJ...859...28Q} {859, 28}

\bibitem[\protect\citeauthoryear{{Riffert} \& {Herold}}{{Riffert} \&
  {Herold}}{1995}]{Riffert-Herold1995}
{Riffert} H.,  {Herold} H.,  1995, \mn@doi [Astrophys. J.] {10.1086/176161},
  \href {http://adsabs.harvard.edu/abs/1995ApJ...450..508R} {450, 508}

\bibitem[\protect\citeauthoryear{{Ripperda}, {Bacchini}  \&
  {Philippov}}{{Ripperda} et~al.}{2020}]{Ripperda-etal2020}
{Ripperda} B.,  {Bacchini} F.,   {Philippov} A.,  2020, arXiv e-prints, \href
  {https://ui.adsabs.harvard.edu/abs/2020arXiv200304330R} {p. arXiv:2003.04330}

\bibitem[\protect\citeauthoryear{{Sano}, {Inutsuka}, {Turner}  \&
  {Stone}}{{Sano} et~al.}{2004}]{Sano-etal2004}
{Sano} T.,  {Inutsuka} S.-i.,  {Turner} N.~J.,   {Stone} J.~M.,  2004, \mn@doi
  [\apj] {10.1086/382184}, \href
  {https://ui.adsabs.harvard.edu/abs/2004ApJ...605..321S} {605, 321}

\bibitem[\protect\citeauthoryear{{Shakura} \& {Sunyaev}}{{Shakura} \&
  {Sunyaev}}{1973}]{Shakura-Sunyaev1973}
{Shakura} N.~I.,  {Sunyaev} R.~A.,  1973, \aap, \href
  {https://ui.adsabs.harvard.edu/abs/1973A&A....24..337S} {500, 33}

\bibitem[\protect\citeauthoryear{{Siegel}, {Ciolfi}, {Harte}  \&
  {Rezzolla}}{{Siegel} et~al.}{2013}]{Siegel-etal2013}
{Siegel} D.~M.,  {Ciolfi} R.,  {Harte} A.~I.,   {Rezzolla} L.,  2013, \mn@doi
  [\prd] {10.1103/PhysRevD.87.121302}, \href
  {https://ui.adsabs.harvard.edu/abs/2013PhRvD..87l1302S} {87, 121302}

\bibitem[\protect\citeauthoryear{{Sironi} \& {Spitkovsky}}{{Sironi} \&
  {Spitkovsky}}{2014}]{Sironi-Spitkovsky2014}
{Sironi} L.,  {Spitkovsky} A.,  2014, \mn@doi [\apjl]
  {10.1088/2041-8205/783/1/L21}, \href
  {https://ui.adsabs.harvard.edu/abs/2014ApJ...783L..21S} {783, L21}

\bibitem[\protect\citeauthoryear{{Sironi}, {Petropoulou}  \&
  {Giannios}}{{Sironi} et~al.}{2015}]{Sironi-etal2015}
{Sironi} L.,  {Petropoulou} M.,   {Giannios} D.,  2015, \mn@doi [\mnras]
  {10.1093/mnras/stv641}, \href
  {https://ui.adsabs.harvard.edu/abs/2015MNRAS.450..183S} {450, 183}

\bibitem[\protect\citeauthoryear{{Stecker}, {Done}, {Salamon}  \&
  {Sommers}}{{Stecker} et~al.}{1991}]{Stecker-etal1991}
{Stecker} F.~W.,  {Done} C.,  {Salamon} M.~H.,   {Sommers} P.,  1991, \mn@doi
  [\prl] {10.1103/PhysRevLett.66.2697}, \href
  {https://ui.adsabs.harvard.edu/abs/1991PhRvL..66.2697S} {66, 2697}

\bibitem[\protect\citeauthoryear{{Svensson} \& {Zdziarski}}{{Svensson} \&
  {Zdziarski}}{1994}]{Svensson-Zdziarski1994}
{Svensson} R.,  {Zdziarski} A.~A.,  1994, \mn@doi [\apj] {10.1086/174934},
  \href {https://ui.adsabs.harvard.edu/abs/1994ApJ...436..599S} {436, 599}

\bibitem[\protect\citeauthoryear{Takahashi}{Takahashi}{2008}]{Takahashi2008}
Takahashi R.,  2008, Monthly Notices of the Royal Astronomical Society, 383,
  1155

\bibitem[\protect\citeauthoryear{{Tchekhovskoy}, {Narayan}  \&
  {McKinney}}{{Tchekhovskoy} et~al.}{2010}]{Tchekhovskoy-etal2010}
{Tchekhovskoy} A.,  {Narayan} R.,   {McKinney} J.~C.,  2010, \mn@doi [\apj]
  {10.1088/0004-637X/711/1/50}, \href
  {https://ui.adsabs.harvard.edu/abs/2010ApJ...711...50T} {711, 50}

\bibitem[\protect\citeauthoryear{{Tchekhovskoy}, {Narayan}  \&
  {McKinney}}{{Tchekhovskoy} et~al.}{2011}]{Tchekhovskoy-etal2011}
{Tchekhovskoy} A.,  {Narayan} R.,   {McKinney} J.~C.,  2011, \mn@doi [\mnras]
  {10.1111/j.1745-3933.2011.01147.x}, \href
  {https://ui.adsabs.harvard.edu/abs/2011MNRAS.418L..79T} {418, L79}

\bibitem[\protect\citeauthoryear{{Thorne}}{{Thorne}}{1974}]{Thorne1974}
{Thorne} K.~S.,  1974, \mn@doi [\apj] {10.1086/152991}, \href
  {https://ui.adsabs.harvard.edu/abs/1974ApJ...191..507T} {191, 507}

\bibitem[\protect\citeauthoryear{{Vourellis}, {Fendt}, {Qian}  \&
  {Noble}}{{Vourellis} et~al.}{2019}]{Vourellis-etal2019}
{Vourellis} C.,  {Fendt} C.,  {Qian} Q.,   {Noble} S.~C.,  2019, \mn@doi [\apj]
  {10.3847/1538-4357/ab32e2}, \href
  {https://ui.adsabs.harvard.edu/abs/2019ApJ...882....2V} {882, 2}

\bibitem[\protect\citeauthoryear{{Xie}, {Lei}, {Zou}, {Wang}, {Wu}  \&
  {Wang}}{{Xie} et~al.}{2012}]{Xie-etal2012}
{Xie} W.,  {Lei} W.-H.,  {Zou} Y.-C.,  {Wang} D.-X.,  {Wu} Q.,   {Wang} J.-Z.,
  2012, \mn@doi [Research in Astronomy and Astrophysics]
  {10.1088/1674-4527/12/7/010}, \href
  {https://ui.adsabs.harvard.edu/abs/2012RAA....12..817X} {12, 817}

\bibitem[\protect\citeauthoryear{{You}, {Straub}, {Czerny}, {Sobolewska},
  {R{\'o}{\.z}a{\'n}ska}, {Bursa}  \& {Dov{\v{c}}iak}}{{You}
  et~al.}{2016}]{You-etal2016}
{You} B.,  {Straub} O.,  {Czerny} B.,  {Sobolewska} M.,  {R{\'o}{\.z}a{\'n}ska}
  A.,  {Bursa} M.,   {Dov{\v{c}}iak} M.,  2016, \mn@doi [\apj]
  {10.3847/0004-637X/821/2/104}, \href
  {https://ui.adsabs.harvard.edu/abs/2016ApJ...821..104Y} {821, 104}

\bibitem[\protect\citeauthoryear{{Younsi}, {Zhidenko}, {Rezzolla}, {Konoplya}
  \& {Mizuno}}{{Younsi} et~al.}{2016}]{Younsi-etal2016}
{Younsi} Z.,  {Zhidenko} A.,  {Rezzolla} L.,  {Konoplya} R.,   {Mizuno} Y.,
  2016, \mn@doi [\prd] {10.1103/PhysRevD.94.084025}, \href
  {https://ui.adsabs.harvard.edu/abs/2016PhRvD..94h4025Y} {94, 084025}

\bibitem[\protect\citeauthoryear{{Zanni}, {Ferrari}, {Rosner}, {Bodo}  \&
  {Massaglia}}{{Zanni} et~al.}{2007}]{Zanni-etal2007}
{Zanni} C.,  {Ferrari} A.,  {Rosner} R.,  {Bodo} G.,   {Massaglia} S.,  2007,
  \mn@doi [\aap] {10.1051/0004-6361:20066400}, \href
  {https://ui.adsabs.harvard.edu/abs/2007A&A...469..811Z} {469, 811}

\bibitem[\protect\citeauthoryear{{Zhang}, {Cui}  \& {Chen}}{{Zhang}
  et~al.}{1997}]{Zhang-etal1997}
{Zhang} S.~N.,  {Cui} W.,   {Chen} W.,  1997, \mn@doi [\apjl] {10.1086/310705},
  \href {https://ui.adsabs.harvard.edu/abs/1997ApJ...482L.155Z} {482, L155}

\bibitem[\protect\citeauthoryear{{Zweibel} \& {Yamada}}{{Zweibel} \&
  {Yamada}}{2016}]{Zweibel-Yamada2016}
{Zweibel} E.~G.,  {Yamada} M.,  2016, \mn@doi [Proceedings of the Royal Society
  of London Series A] {10.1098/rspa.2016.0479}, \href
  {https://ui.adsabs.harvard.edu/abs/2016RSPSA.47260479Z} {472, 20160479}

\makeatother
\end{thebibliography}
\section*{Appendix A: Thin disc setup}
In this study, the initial thin-disc setup is based on the \cite{Novikov-Thorne1973} model, where we use the radiative flux term to model the density profile at the equatorial plane. The scale-height is modeled with \cite{Peitz-Appl1997, Riffert-Herold1995} thin-disc prescription, where the aspect ratio of the accretion disc is considered to be $H/r\ll1$. To test the physical nature of the setup, we consider a hydrodynamical setup (Model HYD) with $A_{r,\theta,\phi}=0$, and all the other parameters are the same as the other models described in section 2.4. In Fig. \ref{fig-19}, we show the normalised density profile for model HYD at simulation time (a) $t=0$, and (b) $t=10000$. We find that the thin-disc structure sufferers minimum changes with temporal evolution (especially the outer part of the disc). It is to be noted that the hydrodynamical model we considered is free from explicit viscous heating and radiative cooling. It is indeed true that radiative cooling plays a crucial role in maintaining the thin-disc structure of the accretion disc. However, due to the local heat balance condition of the thin disc, the combined effect of cooling and explicit viscous heating is negligible (see \cite{Shakura-Sunyaev1973, Paczynski-Kogan1981}, etc.). It, therefore, does not reflect in the dynamical evolution of the model HYD. This makes our setup suitable to study the effect of the magnetic field in the dynamical properties of the accretion disc and corresponding jet launching and disc-wind driving even in the ideal GRMHD limit. 

\begin{figure}
\centering
\includegraphics[scale=0.28]{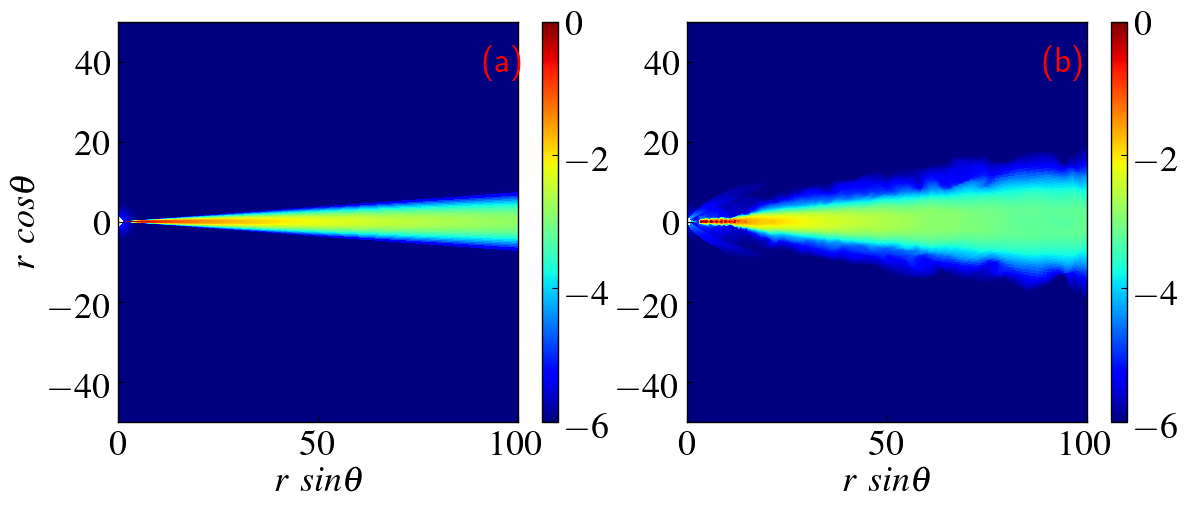}
\caption{Normalised density profile $\rho/\rho_{\rm max}$ for model HYD at time (a) $t=0$ and (b) $t=10000$.}
\label{fig-19}
\end{figure}

\section*{Appendix B: Scale-height test}
It is evident from the equations \ref{eq-06}, \ref{eq-07}, and \ref{eq-09} the scale-height $H$ depends explicitly on $\Gamma$, ${\cal K}$, and $\Theta_0$. In the study we chose $\Gamma=4/3$, ${\cal K}=0.1$, and $\Theta_0=0.001$, with these values the maximum vale of aspect ratio at the outer boundary is $(H/r)_{\rm max}\sim 0.07$. Here, we chose four different models with different values of $\Theta_0=0.005$, $0.001$ (model C), $0.0005$, and $0.0001$, they have $(H/r)_{\rm max}\sim 0.16$, $0.07$, $0.05$, and $0.02$, respectively. All the other parameters are the same as model C.
\begin{figure}
\centering
\includegraphics[scale=0.25]{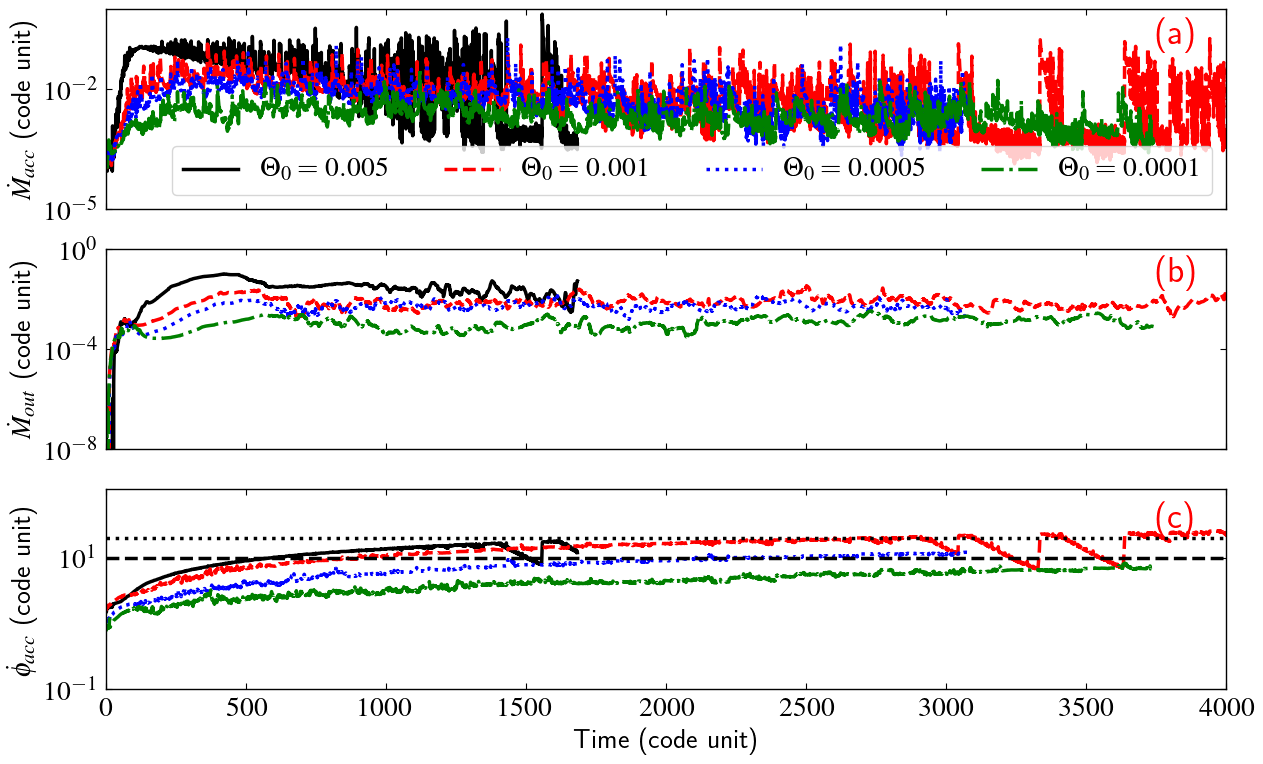}
\caption{Temporal evolution of accretion rate $\dot{M}_{\rm acc}$, outflow rate $\dot{M}_{\rm out}=\dot{M}_{\rm jet} + \dot{M}_{\rm wind}$, and the magnetic flux accumulated at the horizon $\dot{\phi}_{\rm acc}$ for models with different $\Theta_0$ marked on the figure. Dashed and dotted horizontal lines corresponds to $\dot{\phi}_{\rm acc}=10$ and $20$, respectively.}
\label{fig-19}
\end{figure}

We observe that all these models show similar qualitative features as the reference model (model C or $\Theta_0=0.001$), viz. BZ-jet, BP disc-wind, formation of plasmoids, oscillations of the inner part of the accretion disc.  
 However, the initial aspect ratio significantly affects the quantitative features of the accretion-ejection system. In order to study it in detail, in Fig. \ref{fig-19}, we show the temporal evolution of accretion rate $(\dot{M}_{\rm acc})$, outflow rate $(\dot{M}_{\rm out}=\dot{M}_{\rm jet} + \dot{M}_{\rm wind})$, and the magnetic flux accumulated at the horizon $(\dot{\phi}_{\rm acc})$ for models with different aspect ratios. We find that thicker accretion disc shows strong outflows and accretion. Such cases accumulate more magnetic flux at the horizon and become MAD. It makes the inner part of the accretion disc turbulent, and we observe oscillations in the inner part of the accretion disc sooner. On the other hand, thinner accretion discs show weaker outflow and weaker accretion rates. It accumulates magnetic flux slowly and remains in a quasi-steady state for a longer time than a thicker accretion disc.  

\label{lastpage}
\end{document}